\documentclass{jfm}
\usepackage{graphicx}

\usepackage{subcaption}
\usepackage{math_pack}
\usepackage{amsmath}
\usepackage{color}

\shorttitle{Effect of geometry and Reynolds number on the separation behind a bulge}
\shortauthor{J.-P. Mollicone, F. Battista, P. Gualtieri and C.M. Casciola}

\title[Effect of geometry and Reynolds number on the separation behind a bulge]{Effect of geometry and Reynolds number 
on the turbulent separated flow behind a bulge in a channel}

\author{J.-P. Mollicone,
  F. Battista,
  P. Gualtieri
 \and C.M. Casciola\corresp{\email{carlomassimo.casciola@uniroma1.it}}}

\affiliation{Department of Mechanical and Aerospace Engineering, Sapienza University of Rome,
via Eudossiana 18, 00184 Rome, Italy}

\begin{document}

\maketitle

\begin{abstract}
Turbulent flow separation induced by a protuberance on one of the walls of an otherwise 
planar channel is investigated using Direct Numerical Simulations. 
Different bulge geometries and Reynolds numbers -- with the highest friction Reynolds number 
simulation reaching a peak of $\Rey_\tau = 900$ -- are addressed to understand
the effect of the wall curvature and of the Reynolds number on the dynamics of
the recirculating bubble behind the bump. Global quantities reveal that most of the 
drag is due to the form contribution, whilst the friction contribution does not change 
appreciably with respect to an equivalent planar channel flow. The size and position 
of the separation bubble strongly depends on the bump shape and the Reynolds number. 
The most bluff geometry has a larger recirculation region, whilst the 
Reynolds number increase results in a smaller recirculation bubble and a shear 
layer more attached to the bump. The position of the reattachment point  
only depends on the Reynolds number in agreement with experimental data available in the literature. 
Both the mean and the turbulent kinetic energy equations are addressed in such 
non homogeneous conditions revealing a non trivial behaviour of the 
energy fluxes. The energy introduced by the pressure drop follows two 
routes: part of it is transferred towards the walls to be dissipated and part feeds 
the turbulent production hence the velocity fluctuations in the separating shear layer. 
Spatial energy fluxes transfer the kinetic energy into the recirculation bubble
and downstream near the wall where it is ultimately dissipated. 
Consistently, anisotropy concentrates at small scales near the
walls irrespective of the value of the Reynolds number. In the bulk flow and in the 
recirculation bubble, isotropy is restored at small scales and the isotropy recovery 
rate is controlled by the Reynolds number. 
Anisotropy invariant maps are presented, showing 
the difficulty in developing suitable turbulence models to predict separated turbulent flow
dynamics.
Results shed light on the processes of production, transfer and dissipation of 
energy in this relatively complex turbulent flow where non-homogeneous effects 
overwhelm the classical picture of wall bounded turbulent flows which typically 
exploits streamwise homogeneity.

\end{abstract}

\begin{keywords}
%Authors should not enter keywords on the manuscript, as these must be chosen by the author during the online submission process and will then be added during the typesetting process (see http://journals.cambridge.org/data/\linebreak[3]relatedlink/jfm-\linebreak[3]keywords.pdf for the full list)
\end{keywords}

\section{Introduction}
\label{sec:int}

Flow separation consists of fluid flow around bodies becoming detached, causing the 
fluid closest to the object's surface to flow in reverse or different directions, 
most often giving rise to turbulent fluctuations. The flow separation can be induced 
either by geometrical singularities, for example in presence of sharp corners, or by smooth 
geometry variations, as those occurring over a curved wall. The resulting adverse 
pressure gradient is sufficient to  cause flow detachment. 

Given its importance from both theoretical and practical point of view, the study of 
turbulence and separation has long been of interest to the fluid mechanics community, 
see e.g. \cite{Simpson_1989} for a review. However, due to its complexity, this classic 
subject is still widely investigated. The separation of fluid flow from objects 
inevitably results in effects such as increased drag and mixing, momentum and energy 
transfer and vortex shedding. An understanding of such effects is helpful to improve 
road vehicle performance, in the study of fluid-structure interaction, to regulate air 
mixing with other substances such as pollutants or fuels, in the study of boundary 
layer control, see e.g. \cite{Marusic_2014} and \cite{Bai_2014}. In modern 
bioengineering studies  such as in hemodynamics, the nature of the 
flow and the intensity of the shear stresses helps to determine whether lesions 
occur at particular vascular sites, as described by \cite{Epstein_1999}.

Turbulent boundary layers with pressure gradients are a common characteristic of 
many aerodynamic flows such as the flow past airfoils, gas turbine blades, sails 
and diffusers. To correctly predict the behaviour and the efficiency of such 
components, the understanding of separation and reattachment mechanisms together 
with the associated energy behaviour is essential, see e.g.~\cite{harun2013pressure} 
and references therein. The fundamental physics is indeed complex and no entirely 
satisfactory turbulence models for numerical simulation of high Reynolds number 
separated flows are nowadays available. This is mainly due to the complexity of 
the geometries inducing separation and to the difficulty in obtaining sufficiently 
accurate experimental or numerical data for reliable statistical analysis. 

Flow separation occurs in both external and internal flows. In external flows, 
boundary layer separation is induced by strong curvature effects and the associated 
adverse pressure gradient (APG). The understanding of such complex interplay
among flow curvature, APG and separation is considered one of the most challenging  
issues in fluid dynamics both for modelling, \cite{Wilcox_1998}, and most
recent DNSs, \cite{soria2017towards}. In these conditions, the classical scaling of 
turbulent statistics is not valid since the flow separation modifies the Reynolds shear 
stress distribution as discussed by~\cite{Skaare_1994}. In internal flows, such as 
channel or pipe flows, on average the pressure decreases in the flow direction. 
However, the pressure gradient may locally revert due to, for example, an abrupt change 
of section and/or the presence of curved walls. In this case, a localised separated flow occurs 
characterised by a statistically steady recirculation region and by an 
eventual reattachment downstream. In such conditions, turbulence develops in 
highly anisotropic and non-homogeneous conditions. In addition to the non-homogeneous 
effects induced by the wall, it is fundamental to address the non-homogeneous
effects in the streamwise direction where the dynamics of turbulent fluctuations occurs
under rapidly changing conditions, see e.g.~\cite{chen2006scale,gualtieri2010direct} 
for a similar study in the context of turbulent flows subjected to rapid time
variations of the mean flow.

The statistical characterisation of separated flows in presence of
adverse pressure gradients is challenging due to the difficulty to control the actual 
pressure gradient and the ensuing separated flow in presence of curved flows, 
see~\cite{Alam_2000}. The experimental generation of turbulent flow with an APG is 
not standardised and the different approaches employed lead to substantially different 
configurations.
\cite{Skaare_1994} and \cite{Krogstad_1995} performed a detailed study of a turbulent 
boundary layer in presence of a strong APG and constant skin friction coefficient, 
providing a detailed analysis of the turbulence statistics including the budget of the 
kinetic energy. Some studies are focused on the intermediate state of separating and 
re-attaching flow, such as the  experimental study of a boundary layer that is 
maintained on the verge of separation conducted by \cite{Elsberry_2000}. 
On the other hand, \cite{Castro_1998} study the separated flow at the leading edge of a 
flat plate in a wind tunnel considering two different conditions: with and without added 
homogeneous isotropic turbulence. In \cite{Webster_1996} the experimental data of an APG 
boundary layer created by a bump in the wall are provided and a detailed 
analysis of turbulence statistics is discussed. \cite{Dengel_1990} performed 
experimental measurements of an APG turbulent boundary layer reporting different 
cases of pressure distributions, with and without reverse flow, showing the strong 
dependence of the near-wall flow properties on the presence or absence of the 
recirculation region. 
To address the turbulent flow separation on smooth geometry the ERCOFTAC test case 81
has been employed in the literature.
A period hill experiment has been designed by Manhart at TU Munich,~\cite{Rapp_2011}.
This experimental setup is made by nine consecutive 2D bumps to reproduce an infinite 
channel with periodic bumps in the streamwise direction.
\cite{Kahler_2016} carry out several measurements on this experimental setup to address the
separated flow. High resolution particle image  velocimetry and particle tracking velocimetry highlight the crucial role 
of the spatial resolution close to the wall.
As stated by the authors, the difficulties to perform these measurements can be 
compared to those encountered in obtaining reliable  Large Eddy Simulations (LES), see e.g. the discussion in \cite{gualtieri2007preservation}. 

A geometry similar to the ERCOFTAC test case 81 is also employed for validation of
different numerical methods and subgrid models for LES and RANS,  \cite{saric_2007},
\cite{hickel2008implicit}, \cite{peller2006turbulent},  \cite{temmerman2003investigation},  \cite{mellen2000LES}, 
\cite{breuer2009flow}, \cite{diosady2014dns},  \cite{frohlich2005highly}.
From this collection of works, separation and reattachment points or turbulence
intensity in the recirculation bubble are found to strongly depend on modelling and  numerics,
\cite{saric_2007}, \cite{temmerman2003investigation} .

Among the methods used in numerics to introduce an APG, one of the easiest ways is to 
use wall flow suction. Alternatively, the APG can be prescribed by a body force. 
\cite{Na_1998} and \cite{Na_1998_2} performed a Direct Numerical Simulation (DNS) of 
a separated boundary layer on a flat plate using suction and blowing velocity 
distributions at the upper boundary. The inflow condition was taken from Spalart's 
temporally evolving zero pressure gradient (ZPG) simulation, see~\cite{Spalart_1988}. 
\cite{Chong_1998} used these data to analyse the topology of near-wall coherent 
structures using the invariants of the velocity gradient tensor.  The comparison of 
experimental and DNS data is presented in~\cite{Spalart_1993} for turbulent boundary 
layers with different pressure gradients. The DNS was performed using a spectral code 
with a fringe region to deal with periodic conditions in the non-homogeneous streamwise 
direction and the friction velocity at the edge of the boundary layer was prescribed 
to reproduce the pressure gradient of the experiment. A similar numerical technique was 
used by \cite{Skote_1998} for simulations of self-similar turbulent boundary layers 
in adverse pressure gradients prescribed by the freestream velocity.  
\cite{Skote_2002} performed the DNS of separated boundary layer flow with two 
different adverse pressure gradients, while \cite{ohlsson_2010} addressed the 
separation in a three dimensional turbulent diffuser. On to relatively more complex 
geometries, \cite{Le_1997} concentrated on the re-attachment location and 
the skin friction coefficient behind a backward facing step.

Issues related to the separation control have been investigated by \cite{Neumann_2004}, 
by means of Large Eddy Simulations (LES). The LES performed by \cite{Wu_1998} was 
compared to the results by \cite{Webster_1996} and it emerged that the use of a 
coarse resolution with an eddy viscosity model did not allow an accurate description of 
the small coherent vortical structures in the near wall region which were observed in 
experiments. LES has been performed by \cite{Kuban_2012} to evaluate the consistency 
and accuracy with respect to similar DNS simulations. 
Indeed, the sub-grid scale models needed in any LES are expected to hamper the 
physics at the smallest scales, calling for the use of DNS where no modelling 
assumptions are introduced. 
Simulations of channel flow with a lower curved wall were performed 
by \cite{Marquillie_2003} at relatively low Reynolds numbers for a two dimensional case 
to study the onset of nonlinear oscillations. \cite{Marquillie_2008} investigated the 
vorticity and kinetic energy budget downstream of such lower curved wall. 
\cite{Marquillie_2011} and \cite{Laval_2012} expanded on these simulations by studying 
the vorticity and streaks dynamics and linking the streaky structures to the kinetic 
energy production.

The present work deals with the Direct Numerical Simulation (DNS) of a fully turbulent 
channel with a lower curved wall, or bump, which produces the flow separation. 
The simulations are based on the spectral element method, as implemented in Nek5000 \citep{nek5000}. 
The basic domain, a planar channel equipped with a lower curved wall, is essentially 
repeated infinite times and is sufficiently long to allow the flow beyond 
the bump to re-attach. This is numerically obtained with periodic boundary conditions 
in the streamwise direction to avoid artificial inlet and outlet conditions.
The highest Reynolds number simulation reaches 
$\Rey_\tau=900$ over the bump that is, presumably, one of the highest friction Reynolds 
number achieved for such a configuration.

%%%%%%%%%%%%%%%%%%%%%%%%%%%%%%%%%%%%%%%%%%
% Added to "describe aims of paper" and to reply to "nothing new" comment
%%%%%%%%%%%%%%%%%%%%%%%%%%%%%%%%%%%%%%%%%%

The objective is to study the effects of the bump geometry and Reynolds number 
on flow separation. One of the global quantities available from experiments is the position of the reattachment point, an elusive quantity to reproduce in numerical simulations, due to the need of using closure models to reach  
sufficiently high Reynolds numbers. In our case,  we can directly compare the simulations with the experiments, observing a good agreement with the available data.
Beside classical first and second-order statistics, the present DNSs provide access to high quality data concerning pressure and friction drag  and to wall shear stress and pressure coefficient distributions at the walls.
In the present flow geometry, the energetics of the flow is rather complex and needs an accurate discussion.
In particular,  the shear layer at the boundary of the separation bubble acts as source of turbulent kinetic energy, 
which is spatially redistributed through the domain by the associated energy fluxes.
In the analysis a crucial role is played by the corresponding terms in the kinetic energy budget of the mean flow,
which are usually trivial in absence of separation.
Locally the flow turns out to be strongly anisotropic, with anisotropy persisting down to the smallest scales.
This effect was already discussed for the zero pressure gradient boundary layer, \cite{jacob2008scaling}, and for the homogeneous shear flow,  \cite{casciola2007residual}. In the present case, the analysis of the anisotropy 
of both large and small scales is studied via the deviatoric components of the Reynolds stresses and the pseudo-dissipation tensor. 
Increasing the Reynolds number, isotropy recovery at small scale is found to occur in the recirculating bubble. However, the anisotropy persists in the shear layer where the  production of turbulent kinetic energy overwhelms the energy cascade forcing the shear scale to approach the dissipative scales.
The anisotropy invariant maps of the Reynolds stresses are finally used to quantify 
the different anisotropic states of the large turbulent scales. 
The results confirm that the present flow poses a significant challenge for turbulence modelling due to the existence of the recirculating bubble behind  the bump and the adverse pressure gradient region along the opposite wall.

The paper is organised as follows: the dataset  is presented in section \S\ref{sec:sim} together with some basic
statistics used for validation. The main results are reported in section  \S \ref{sec:res} which is divided into several subsections illustrating  different topics, i.e. instantaneous flow fields, Reynolds stress tensor, budget of mean and 
turbulent kinetic energies and anisotropy analysis, including anisotropy invariant maps.
The last section \S \ref{sec:final} summarises the main findings of the paper.

\section{Simulations}
\label{sec:sim}

\begin{table}
\centering
\begin{tabular}{ cccccccc } 
\hline
 Simulation   &   $\Rey$   &   $ \Rey_{\tau}  $  &   $ \overline{\Rey_{\tau}} $  &   $ \Delta {x^+} $   &   $ \Delta {z^+} $   &   $ \Delta {y^+}_{max/min} $  &  $ a $  \\  
 \hline
 A1 & 2500   & 300 &  160 & 2.8  & 2.8  & 3.7/0.5  &  0.15  \\
 B1 & 2500   & 300 &  160 & 2.8  & 2.8  & 3.7/0.5  &  0.25  \\
 C1 & 2500   & 300 &  160 & 2.8  & 2.8  & 3.7/0.5  &  0.50  \\
 A2 & 5000   & 550 &   280 & 4.4  & 5.0  & 6.0/0.7  &  0.15  \\
 A3 & 10000 & 900 &   550 & 6.5  & 7.0  & 9.5/0.9  &  0.15  \\
 \hline
\end{tabular}
\caption{Simulation matrix. The nominal Reynolds numbers is $\Rey=h_0 U_b/\nu$ 
where $h_0$ and
$U_b$ are the half nominal channel height and the bulk velocity respectively. 
$\Rey_{\tau}=h u_\tau/\nu$ is the maximum friction Reynolds number taken at the 
bump tip with $u_\tau = \sqrt{\tau_w/\rho}$ the local shear velocity 
($\tau_w $ is the local mean shear stress and $\rho$ is the density), and $h$ half 
the local channel height. The average friction Reynolds number is
denoted with $\overline{\Rey_{\tau}}$ where averages are performed on both the 
upper and lower walls. $\Delta {x^+} $, $ \Delta {z^+} $  and  
$ \Delta {y^+}_{max/min} $  are the spatial resolution in the streamwise, spanwise and 
wall-normal directions made dimensionless with the average wall-unit.
The parameter $a$ determines the different bump geometries, see text.}
\label{tab:simgrid}
\end{table}
\begin{figure}
  \centerline{\includegraphics[width=0.9\textwidth]{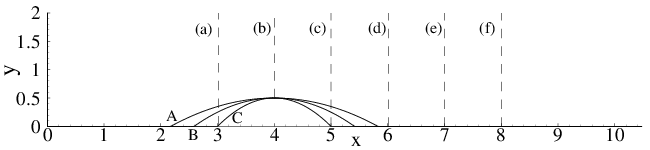}}
\caption{Sketch of the different bump geometries given by $y=-a\,(x-4)^2+0.5$ where
$a$ is reported in table~\ref{tab:simgrid} and localisation of the different stations
where statistics are addressed.
\label{fig:stations}}
\end{figure}

\begin{figure}
\includegraphics[width=\textwidth]{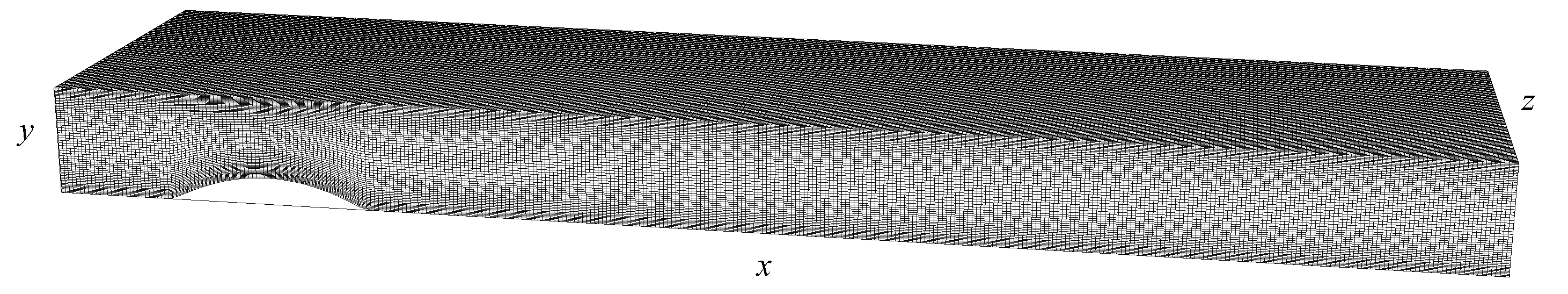}
\caption{\label{fig:geom} Sketch of the geometry of the channel with the curved lower 
wall for simulation (A). Periodic conditions are enforced in the streamwise, $x$, and 
spanwise, $z$, direction. No slip and impermeability are enforced on the top and 
bottom walls.}
\end{figure}

Five different simulations, whose parameters are summarised in table~\ref{tab:simgrid}, 
have been carried out.  Simulations A1, B1 and C1 have the same Reynolds number but 
different bump geometries, going from the most streamlined (A1) to the most bluff (C1) 
profile, see figure~\ref{fig:stations}. A sketch of the whole three dimensional domain 
is shown in figure \ref{fig:geom}. Simulations A2 and A3 have the same bump geometry 
as simulation A1 but are performed at higher Reynolds numbers. 
Table~\ref{tab:simgrid} lists the nominal Reynolds number $Re$, the maximum 
friction Reynolds number $Re_{\tau}$ on the bump, the friction Reynolds number 
averaged on the top and bottom walls $ \overline{\Rey_{\tau}}$ and the grid 
spacing in all directions. 
The grid is uniform in the streamwise and spanwise directions, and stretched in the wall normal direction to cluster grid nodes toward the walls, see table~\ref{tab:simgrid}.
The nominal Reynolds, $\Rey=h_0 U_b/\nu$, is defined in terms 
of half the channel height, $h_0$, of the bulk velocity, $U_b$, and of 
the kinematic viscosity, $\nu$. The friction Reynolds number is 
$\Rey_{\tau}=h u_\tau/\nu$, with $u_\tau = \sqrt{\tau_w/\rho}$ the shear velocity 
($\tau_w $ is the local mean shear stress and $\rho$ is the density), and $h$ half 
the local channel height. DNS is employed to solve the incompressible Navier-Stokes 
equations, 
\begin{equation}
\pd{u_i}{t} + u_j \pd{u_i}{x_j} = - \pd{p}{x_i} + \frac{1}{Re} \pd{^2u_i}{x_j^2} 
\quad\quad\quad \pd{u_i}{x_i}=0,
\label{eq:NS}
\end{equation}
where $u_i$ is the $i^{th}$ velocity component, and $p$ is the hydrodynamic pressure. 
Henceforth all length scales are made dimensionless with the nominal channel 
half-height, $h_0$, time scales with $h_0/U_b$ and pressures with $\rho U_b^2$. 

The simulations are carried out using Nek5000, see~\cite{nek5000}, which is an 
open-source code that can simulate unsteady incompressible and low Mach number flows.
The discretisation is based on the Spectral Element Method (SEM), 
see~\cite{Patera_1984}, whose formulation allows for Direct Numerical Simulations. 
Highly accurate numerical approaches for the simulation of wall bounded turbulent 
flows are crucial since it is desirable that the numerical error does not contaminate 
the multi-scale non-linear interactions. This feature is fulfilled by the SEM 
approach, which reconciles the high accuracy, typical of a spectral method, and 
the flexibility (in terms of geometrical configuration), typical of finite element 
approaches.

\begin{figure}
  \centerline{ \includegraphics[width=0.75\textwidth]{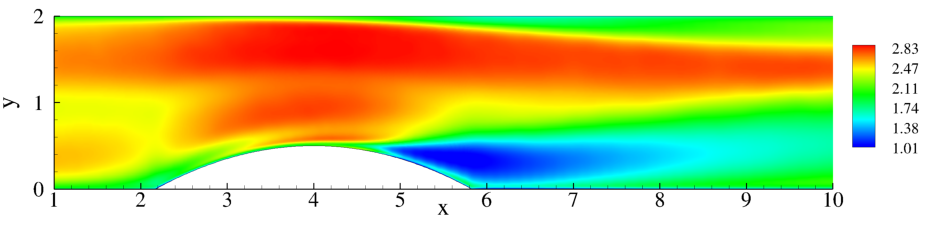}}
 \caption{Kolmogorov scale with respect to the local grid spacing, 
 $ \pi \eta /  \Delta $, for simulation A3, where $\Delta=\sqrt[3]{\Delta x \Delta y \Delta z}$. }
\label{fig:eta_deltay}
\end{figure}

The grid spacing in the wall normal direction $y$ at the centre of the domain and at the 
walls is given by $ \Delta {y^+}_{max} $ and $ \Delta {y^+}_{min} $, respectively. 
The superscript $+$ denotes wall units referred to the average friction Reynolds number. 
The uniform grid spacing in the streamwise $x$ and spanwise direction $z$ in 
inner units is denoted by $ \Delta {x^+} $ and $ \Delta {z^+} $ respectively.
These values, reported in table~\ref{tab:simgrid}, are well within the grid resolution suggested 
by~\cite{Kim_1987} for well resolved DNS of wall bounded turbulent flows.
Comparison of the local grid spacing, $\Delta=\sqrt[3]{\Delta x \Delta y \Delta z}$, 
with the local Kolmogorov scale, 
$\eta=\left(\nu^3/\epsilon_T \right)^{1/4}$ where $\epsilon_T$ is the turbulent
kinetic energy dissipation rate, is shown in figure~\ref{fig:eta_deltay},
in particular the quantity $\pi \eta /\Delta$ is reported for the highest Reynolds number case.
The resolution requirement for a classical spectral method is $k_{max}\eta =\pi \eta /\Delta >1$. 
In the present case $\pi \eta /\Delta$ ranges between $1$ in the 
recirculation bubble and almost $3$ in the bulk of the flow, values 
adequate for the high fidelity reconstruction of the small scale 
dynamics of the flow, given the accurate dispersion characteristics of
the spectral element method.

The bump profile on the lower wall is generated using the  equation $y=-a\,(x-4)^2+0.5$, where the 
coefficient $a$ is reported in  table~\ref{tab:simgrid} for each simulation. The mesh for the 
lower Reynolds number case contains approximately 120 million grid points and the 
simulation was run on 8192 cores, using approximately 6 million core hours. 
The mesh for the higher Reynolds number case was run with approximately 400 million 
grid points on 32768 cores using approximately 30 million core hours. All simulations 
were run with a spectral element order $N=9$ except for the high Reynolds number simulation 
(A3) which was run at a spectral element order $N=11$. 
The reason for changing the spectral order is purely technical, motivated by the need of optimising the machine performance at changing dimensions of the simulation. All simulations were run on the FERMI Blue Gene/Q Tier0 system at the CINECA supercomputer centre in Bologna, Italy. 

The geometry is shown in figure~\ref{fig:geom}. The domain has dimensions 
$(L_x \times L_y \times L_z) = (26 \times 2 \times 2\pi )$ to avoid flow 
confinement at high Reynolds number, see~\cite{lozano2014effect} for similar 
issues in the context of planar channel flows.
In the pictures, the flow is from left to right in the $x$ direction with periodic boundary conditions in 
both the $x$ and $z$ directions. No slip and zero normal velocity boundary conditions 
are imposed at the top and bottom walls. Accounting for periodicity, the actual geometry consists of an infinite channel  with periodic bumps in the streamwise direction that are spaced by approximately 44
bump heights. 
The distance between consecutive bumps is enough to allow flow reattachment and to minimise the effects 
that the separation behind the fore bump may have on the aft bump. In this way,
the use of inflow conditions, either synthetic or  provided by companion channel simulations, is avoided.  The flow is sustained by an overall  pressure drop $\Delta p(t)$ in the $x$ direction that is modulated in time to keep 
the same constant flow rate for all simulations.

% Temporal and spatial correlations
\begin{figure}
  \centerline{
  \includegraphics[width=0.32\textwidth]{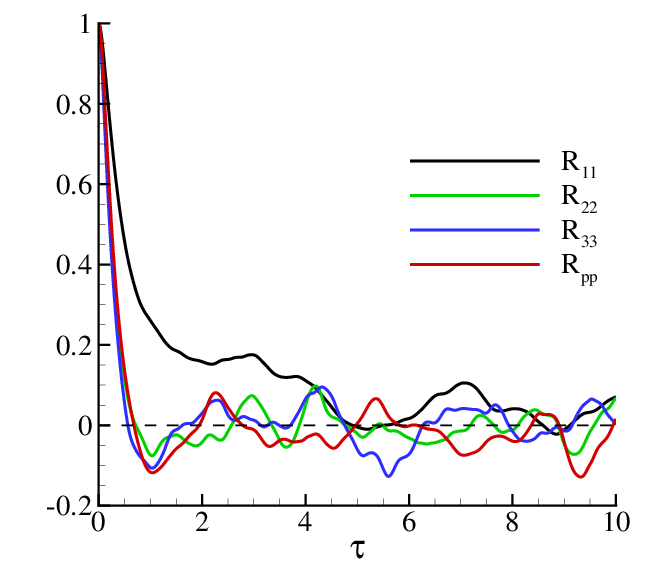}
    {\scriptsize \put(-125,98){\bf (a)}}
  \includegraphics[width=0.32\textwidth]{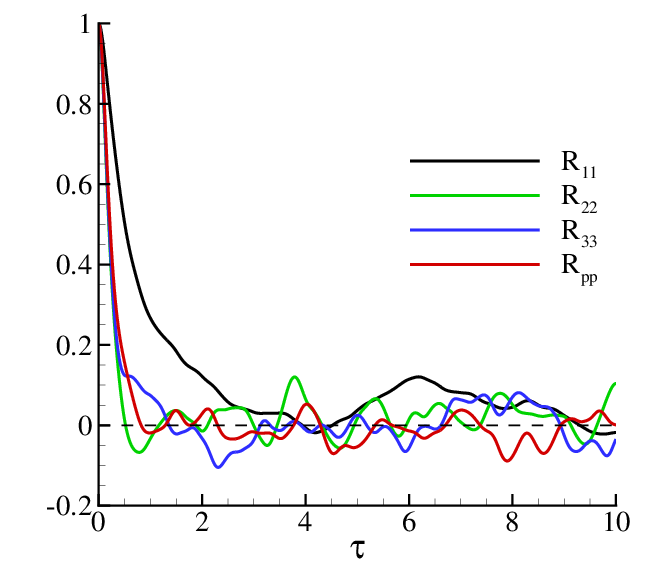}
    {\scriptsize \put(-125,98){\bf (b)}}
  \includegraphics[width=0.32\textwidth]{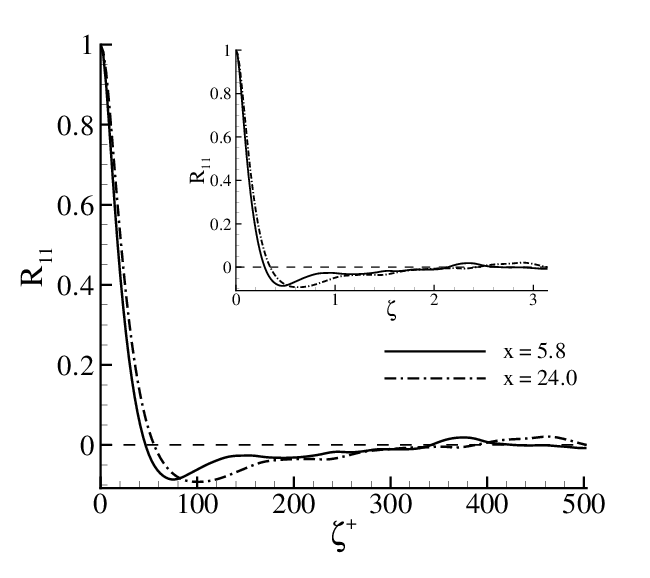}
    {\scriptsize \put(-125,98){\bf (c)}}}
  \caption{Temporal auto-correlation of velocity and pressure signals, for simulation A1, 
   probed just after the bump at $x=5.8$ in panel (a) and at $x=24$ in panel (b). 
   Panel (c) shows the spatial correlation of axial velocity fluctuations in the spanwise 
   direction at $x=5.8$ and at $x=24$. All panels refer to the same distance from the upper 
   planar wall, $d = 2- y=1.6$. $\tau$ is normalised with $h_0/U_b$.}
  \label{fig:correlation}
\end{figure}
Approximately 500 statistically uncorrelated fields, separated by a time interval of 
$\Delta t_{stat}=6$, were collected for each simulation in order to obtain properly 
converging statistics. 
$\Delta t_{stat}$ is normalised with $h_0/U_b$.
Defining the ``flow-through time'',   $t_{ft}$, as the time  
needed for a turbulent structure to travel all along the channel length, 
see~\cite{Kahler_2016}, the simulation time is $T_{tot}= 3000 \simeq 115 \, t_{ft}$, 
which makes sure that the velocity statistics converge~\citep{Kahler_2016}. 
Figure~\ref{fig:correlation} shows, for simulation A1, the temporal auto-correlation of velocity,
\begin{align*}
R_{ii}(x,y,\tau) & =\lim_{T \rightarrow \infty }  \frac{1}{\pi T {u_i}_{\rm rms}^2}\int_0^T 
\int_0^\pi u'_i(x,y,z,t) u'_i(x,y,z,t+\tau) dz dt \\ 
& = \frac{\langle u'_i(x,y,z,t) u'_i(x,y,z,t+\tau) \rangle}{{u_i}_{\rm rms}^2}
\, , \qquad \mbox{no sum},
\label{eq:tcorr}
\end{align*}
\begin{figure}
  \centerline{
  \includegraphics[width=0.32\textwidth]{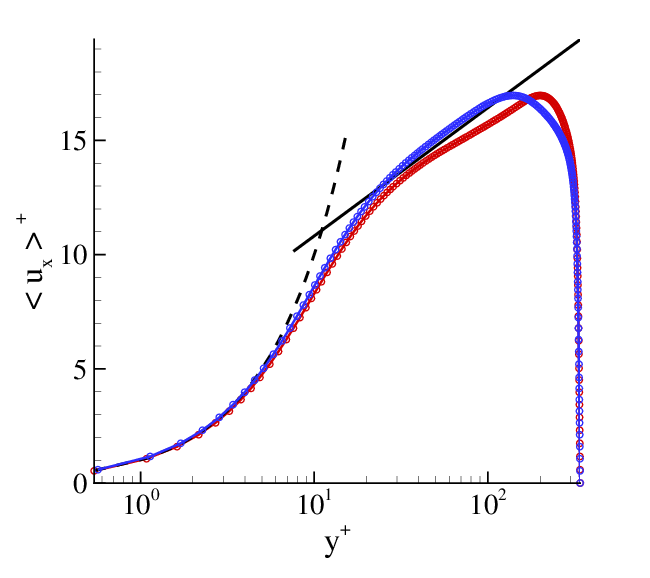}
    {\scriptsize \put(-125,98){\bf (a)}}
  \includegraphics[width=0.32\textwidth]{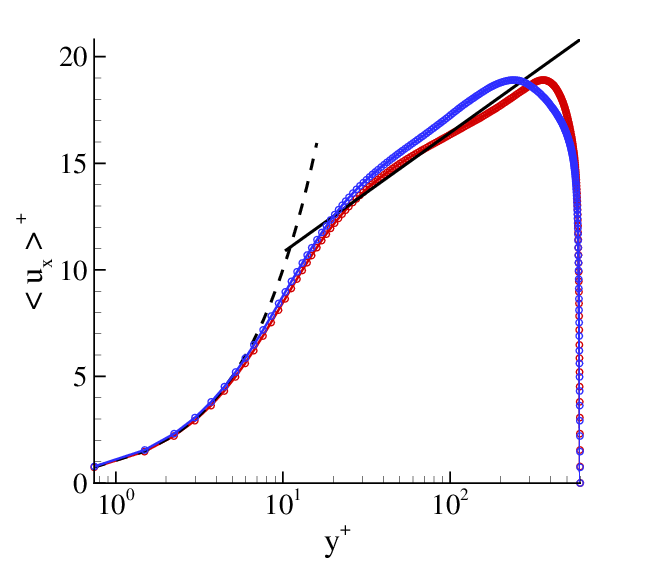}
    {\scriptsize \put(-125,98){\bf (b)}}
  \includegraphics[width=0.32\textwidth]{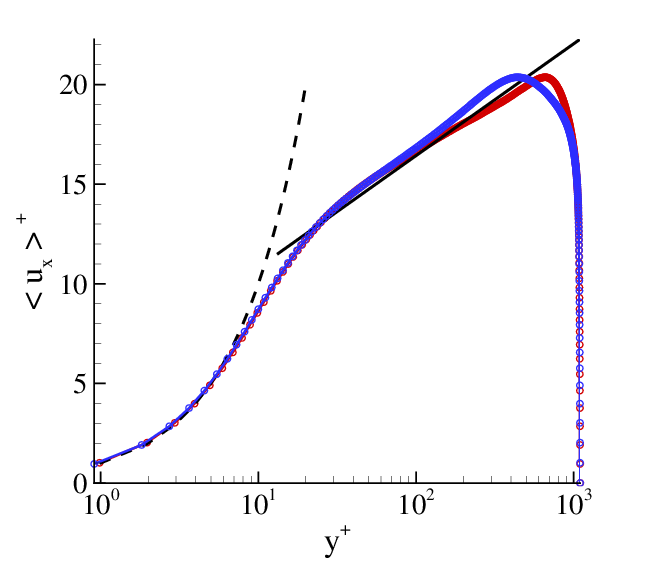}
    {\scriptsize \put(-125,98){\bf (c)}}}
 \caption{\label{fig:log_law} Plot of mean stream-wise (x-direction) velocity normalised with 
  friction velocity, $\langle u_x\rangle^+=\langle u_x\rangle/u_\tau$, against $y^+$ at 
  $x=24$ for simulations A1, A2 and A3 in panels (a), (b) and (c), respectively. The top and 
  bottom wall velocities are represented by the blue and red lines, respectively. 
  The dashed black line is the theoretical prediction, $\langle u_x\rangle^+=y^+$, in the 
  viscous sub-layer. The solid black line is the theoretical prediction, 
  $\langle u_x\rangle^+=1/k \, \log{(y^+)}+A$, in the log-layer region with 
  $k=0.41$ and $A=5.$}
\end{figure}
and pressure signals, 
$R_{pp}(x,y,\tau) = \langle  p'(x,y,z,t) p'(x,y,z,t+\tau)\rangle/p_{\rm rms}^2$, 
probed at different locations inside the domain with the local root mean square 
fluctuation defined to normalise to one the correlation at zero time separation, i.e. 
$R_{ii}(x,y, 0) = 1$. Angular brackets indicate averages over the 
homogeneous coordinates, $z$ and $t$, while a prime indicates the fluctuation with 
respect to the local mean value. Some probes are located just beyond 
the bump (panel (a) of figure~\ref{fig:correlation}) and some others in the fully 
reattached flow (panel (b) of figure~\ref{fig:correlation}). The correlations confirm 
that fields separated by $\Delta t_{stat}$ are uncorrelated. For the same points, 
the spatial correlation of streamwise velocity fluctuations, 
$R_{xx}(x,y,\zeta) = \langle u'_x(x,y,z,t) u'_x(x,y,z+\zeta,t) \rangle/{u_x}_{\rm rms}^2$,
in the spanwise direction, $z$, is shown in panel (c) of figure~\ref{fig:correlation}. 
The solid line refers to the correlation just beyond the bump whilst the dashed line 
refers to the correlation in the fully reattached flow. The spatial separation is 
normalised with (nominal) wall units $\zeta^+=\zeta \overline{\Rey_\tau}$. In the 
reattached flow region, the minimum correlation occurs at $\zeta^+\simeq100$ assuring 
that the spanwise length is suitable to avoid confinement effects on the turbulent 
structures. Close to the bump, the correlation minimum occurs at $\zeta^+\simeq80$. 
The inset in the figure reports the same quantities as a function of the spanwise 
separation normalised with the external unit.

Figure~\ref{fig:log_law} shows plots of dimensionless mean stream-wise (x-direction) 
velocity, $\langle u_x\rangle^+ = \langle u_x\rangle/u_\tau$, against 
$y^+=y\,\overline{\Rey_\tau}$ at $x=24$ averaged in the top half of the channel (blue) and  
in the bottom half (red), for simulations A1, A2 and A3. At this station, the flow almost 
entirely recovers the structure of a canonical turbulent channel flow. The plots show a 
well resolved viscous sub-layer, the buffer layer and the expected log-law region. 
The theoretical prediction for the viscous region close to the wall is represented by 
the dashed black line, $\langle u_x\rangle^+ = y^+$.
The symbols in the plots denote actual data points, showing the high 
resolution achieved in the simulation. The solid black line represents the log-law,  
$\langle u_x\rangle^+=1/k \log{y^+}+A$. The figure shows that, both at the bottom and top 
wall, this law is approached by the present data with better accuracy as the Reynolds 
number is increased, see the caption of the figure for the values of Von-Karman constant 
and intercept which are in agreement with those found in, 
e.g., \cite{nagib2008variations,marusic2013logarithmic}.

\section{Results}
\label{sec:res}

\subsection{Instantaneous flow fields}
\label{sec:inst}

% Instantaneous xy planes
\begin{figure}
  \centerline{\includegraphics[width=\linewidth]{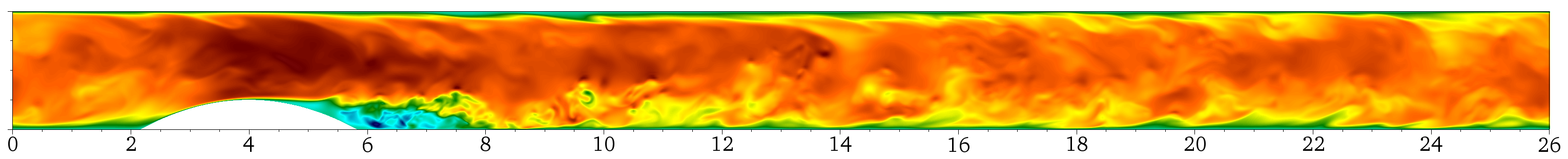}
  {\scriptsize \put(-380,42){\bf (A1)}}}
  \centerline{\includegraphics[width=\linewidth]{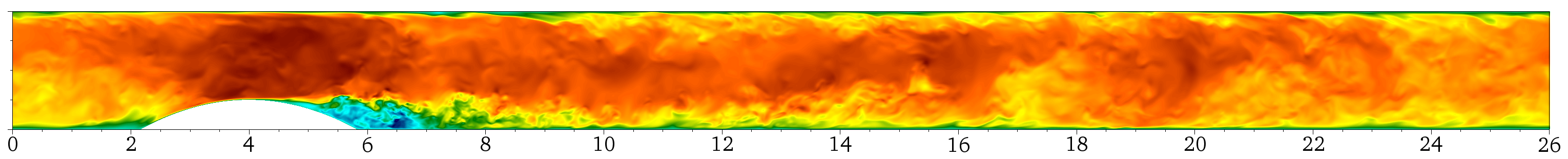}
  {\scriptsize \put(-380,42){\bf (A2)}}}
  \centerline{\includegraphics[width=\linewidth]{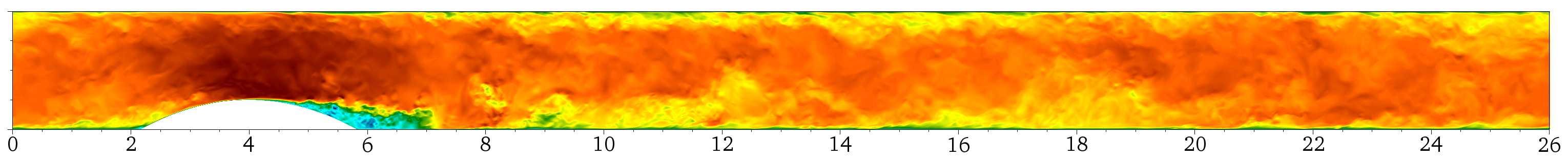}
  {\scriptsize \put(-380,42){\bf (A3)}}}
  \centerline{\includegraphics[width=\linewidth]{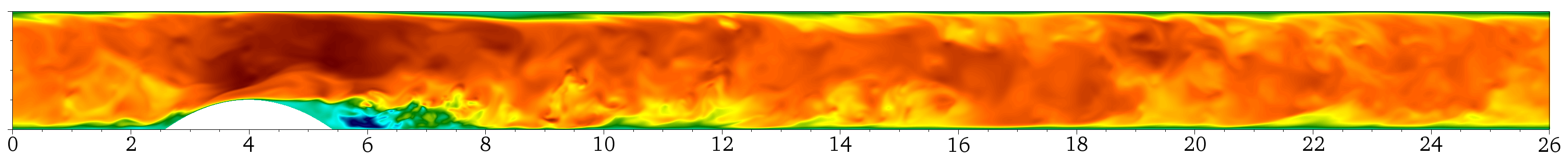}
  {\scriptsize \put(-380,42){\bf (B1)}}}
  \centerline{\includegraphics[width=\linewidth]{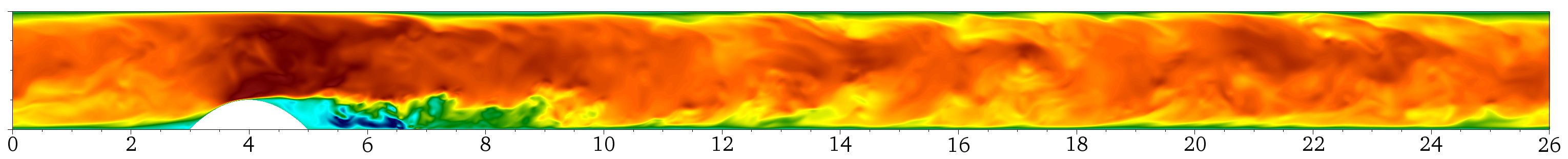}
  {\scriptsize \put(-380,42){\bf (C1)}}}
  \centerline{\includegraphics[width=0.3\linewidth]{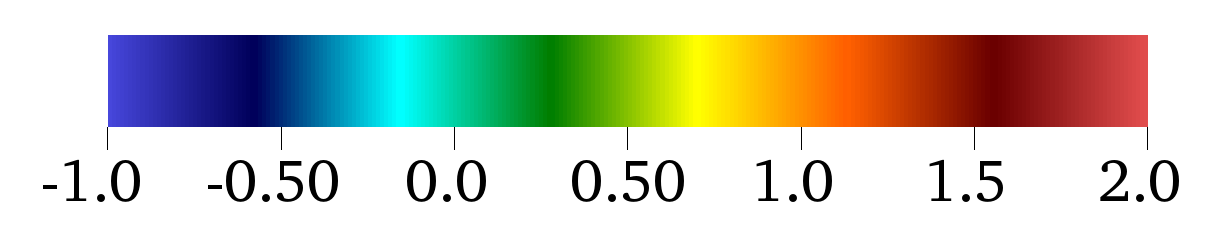}   }
  \caption{Instantaneous stream-wise velocity contour plots in x-y planes for all five simulations.}
\label{fig:inst_xvel}
\end{figure}
%

% A1 and A3 - in pairs - zx plane
\begin{figure} 
  \centerline{\includegraphics[width=0.8\linewidth]{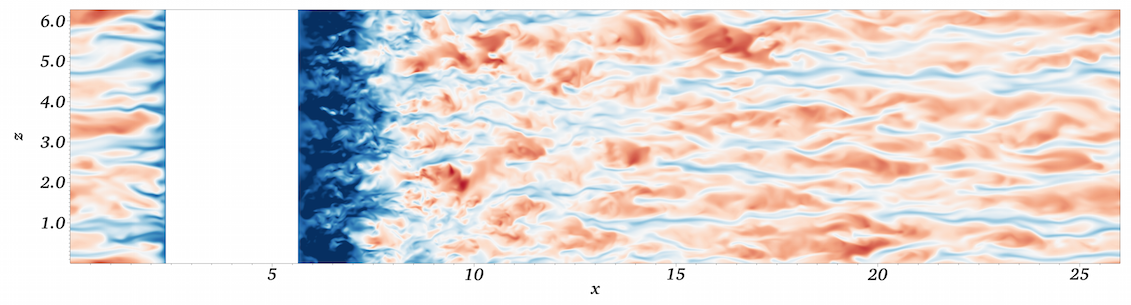}
  {\scriptsize \put(-325,76){ \bf (a)}}}
  \centerline{\includegraphics[width=0.8\linewidth]{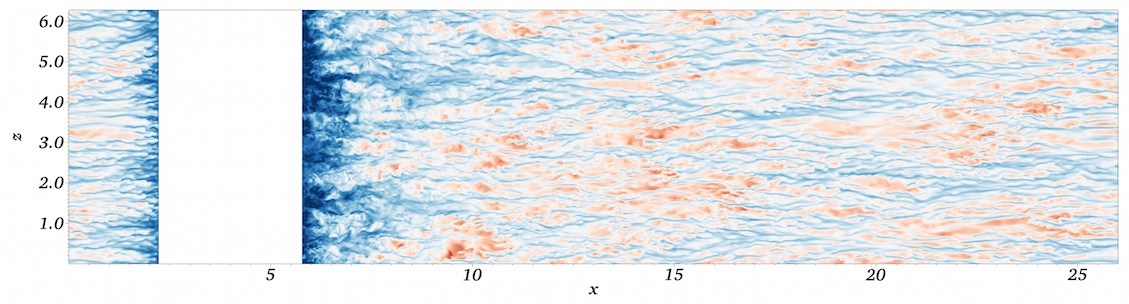}
  {\scriptsize \put(-325,76){ \bf (b)}}}
  \centerline{\includegraphics[width=0.8\linewidth]{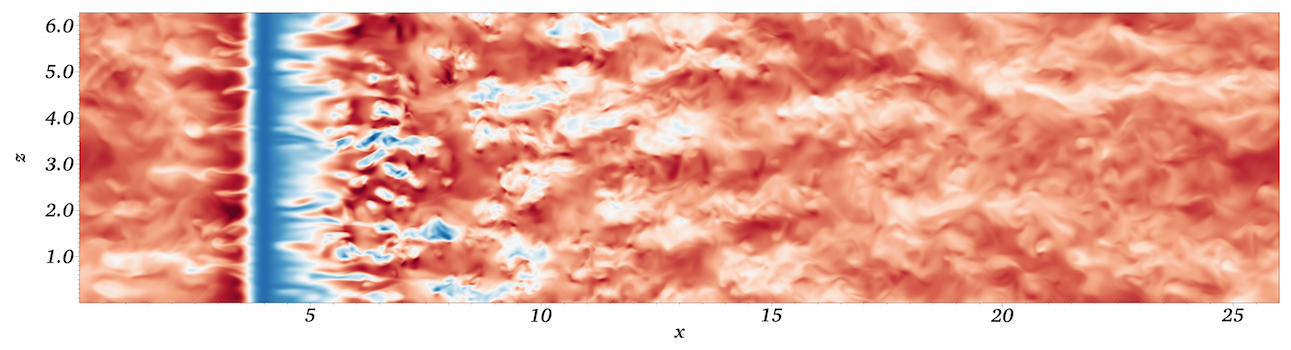}
  {\scriptsize \put(-325,76){ \bf (c)}}}
  \centerline{\includegraphics[width=0.8\linewidth]{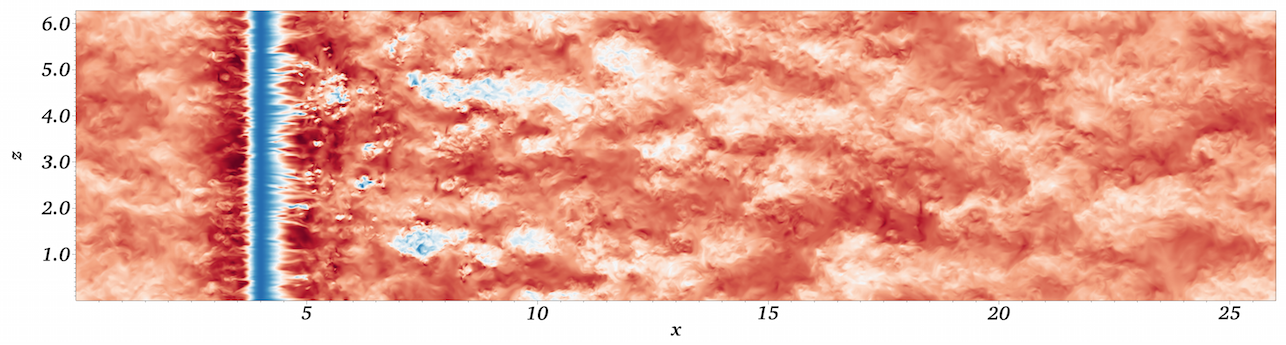}
  {\scriptsize \put(-325,76){ \bf (d)}}}
  \centerline{\includegraphics[width=0.8\linewidth]{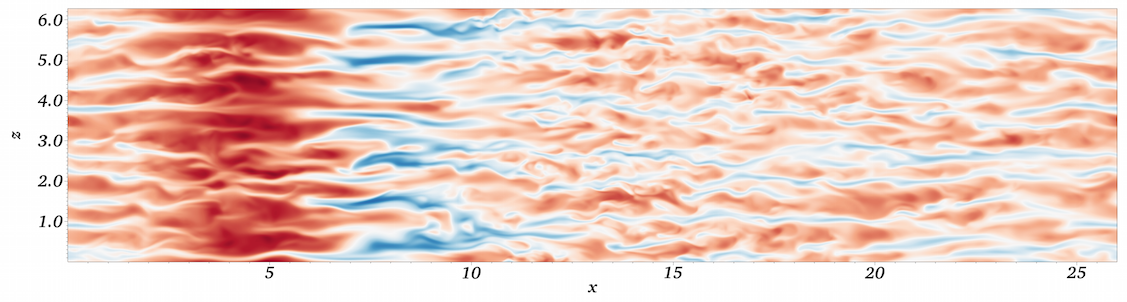}
  {\scriptsize \put(-325,76){ \bf (e)}}}
  \centerline{\includegraphics[width=0.8\linewidth]{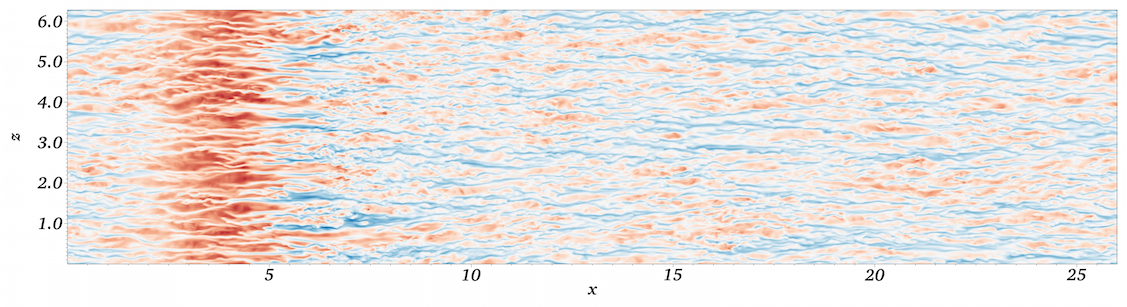}
  {\scriptsize \put(-325,76){ \bf (f)}}}
  \centerline{\includegraphics[width=0.3\linewidth]{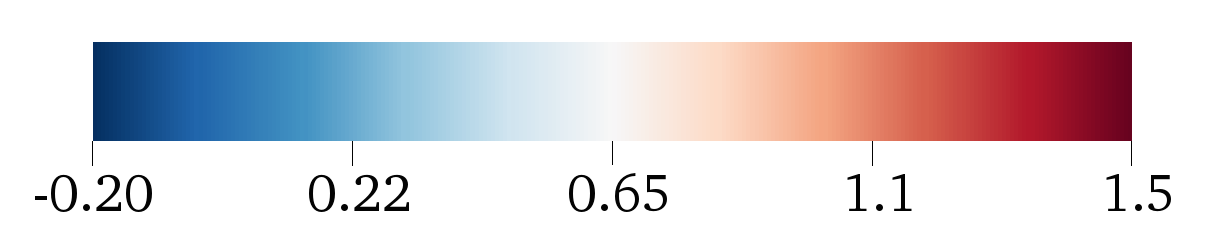}   }
  \caption{Instantaneous stream-wise velocity contour plots in x-z planes at $y^+ = 15$ in panels (a) and (b), 
  $y=0.5$ in panels (c) and (d) and $y^+=15$ (from the top wall) in panels (e) and (f). Simulation A1 in panels (a), (c), (e) and simulation A3 in 
  panels (b), (d) and (f). }
\label{fig:r13_zx}
\end{figure}

Figure~\ref{fig:inst_xvel} shows instantaneous contour plots of streamwise velocity 
in the x-y plane for all the simulations. As expected, investing the bump, the flow velocity
increases at the channel restriction and separates behind the bump with 
the formation of an intense shear layer between the bulk flow  and the separation bubble close to 
the bottom wall. With increasing Reynolds number, cases A1, A2 and A3 progressively, structures at 
smaller scales appear. The separated region behind the bump becomes smaller, 
more attached to the bump and less protruded in the streamwise direction.
It is interesting to compare, at least qualitatively, the flow
snapshots for cases A1 and A3 with the smoke patterns used to visualise the boundary 
layer on a convex wall as shown in pg.~91 of the classical album 
by~\cite{Vandyke_1982}. Indeed, in case A3 the flow impinging the bump is clearly 
characterised by small scale structures while case A1, even though nominally turbulent,
appears smoother. Under this respect, case A3 can be regarded as producing a turbulent 
boundary layer between the external turbulent stream and the wall able to better withstand 
the adverse pressure gradients before separation. The separated region in front of the bump is also smaller at high Reynolds number. 
On the other hand, the separated region behind the bump becomes larger as 
the bump becomes bluffer. At the top wall, the boundary layer thickens after the 
constriction due to the adverse pressure gradient but separation is not observed.
This effect is more evident for the lower Reynolds number cases, probably due to 
the higher extension of the recirculation bubble.

Figure~\ref{fig:r13_zx} shows instantaneous contour plots of streamwise velocity in x-z 
planes at three wall normal distances for simulations A1 and A3 to qualitatively compare 
Reynolds number effects. At the higher Reynolds number, smaller structures are 
clearly present and the spacing between region of high and low speed is greatly 
reduced. In the far-downstream region and close to the walls, this is consistent with the 
expected scaling of streak spacing in internal units, \cite{robinson1991coherent}. In 
particular, panels (a) and (b) show xz-planes close to the bottom wall at $y^+=15$.
This wall distance corresponds to the classical buffer layer of the channel flow. 
The empty region in the plot represents the intersection of the plane with the bump. 
The recirculation region behind the bump is characterised, at this distance from the wall, 
by negative velocity and small-scale structures. The size of the region 
where reverse flow occurs decreases when increasing the Reynolds 
number. Downstream, the small scale structures elongate in the flow direction and resemble 
the streaky structures found in turbulent planar channel flow. 
The xz-planes for simulations A1 and A3, in panels (c) and (d), respectively, just touch 
the top of the bump which is indicated by the continuous zero velocity line 
at $x=4$ in the contour plot. The acceleration of the flow just before the bump and its 
deceleration just behind can be observed. The small scale structures far 
away from the bump increase their length downstream. For simulation A3, the flow structures 
generated in the shear layer appear confined in a smaller region since the flow is more 
attached to the bump's surface and the separated region protrudes less in the 
streamwise direction. Panels (e) and (f) show the xz-planes close to the top wall, at $y^+=15$. 
The effect of the bump on the velocity is still present and a clear velocity 
increase is seen at the bump's location, due to the cross-section restriction. This 
is followed by a low velocity region corresponding to the end of the separation bubble 
at the opposite wall. In this region, the turbulent structures maintain a streamlined, 
streaky shape and no separated region is present.

\subsection{Mean velocity and turbulent fluctuations }

\begin{figure}
  \centerline{\includegraphics[width=0.7\linewidth]{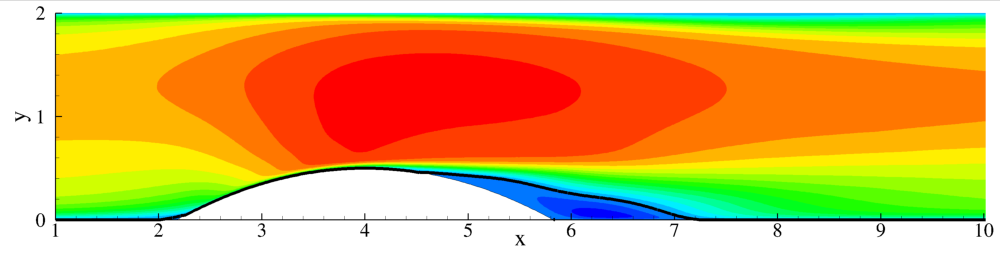}
  {\scriptsize \put(-300,63){\bf (A1)}}}
  \centerline{\includegraphics[width=0.7\linewidth]{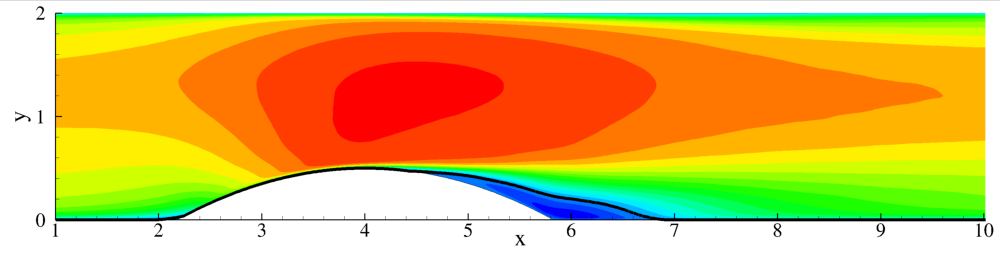}
  {\scriptsize \put(-300,63){\bf (A2)}}}
  \centerline{\includegraphics[width=0.7\linewidth]{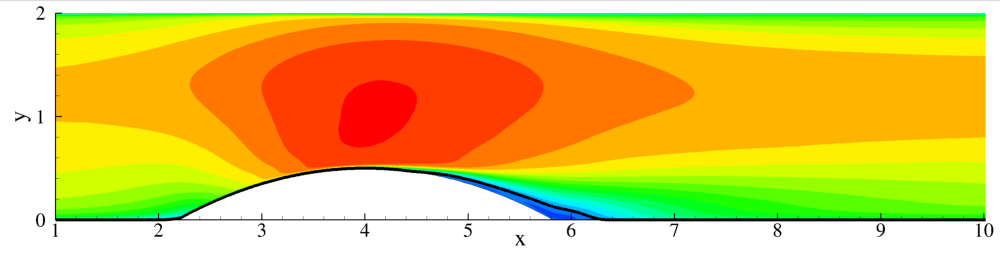}
  {\scriptsize \put(-300,63){\bf (A3)}}}
  \centerline{\includegraphics[width=0.7\linewidth]{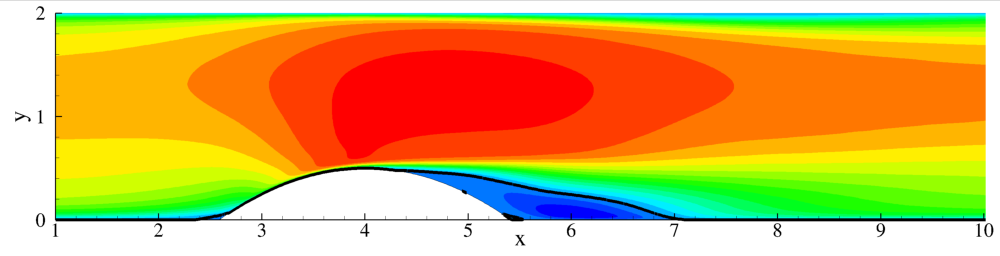}
  {\scriptsize \put(-300,63){\bf (B1)}}}
  \centerline{\includegraphics[width=0.7\linewidth]{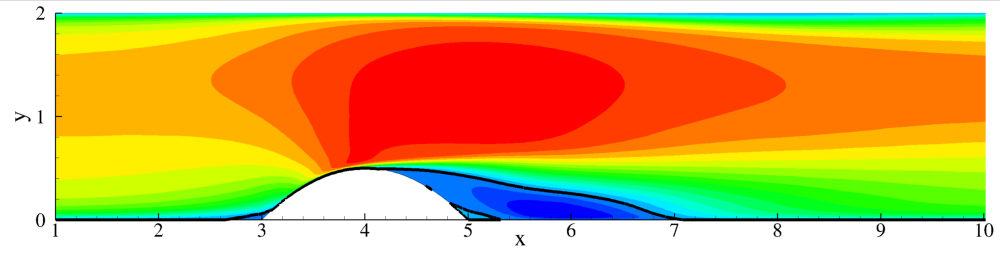}
  {\scriptsize \put(-300,63){\bf (C1)}}}
  \centerline{\includegraphics[width=0.3\linewidth]{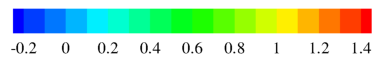}       }
  \caption{Average streamwise velocity $\langle u_x \rangle$ contour plots with isoline at $\langle u_x \rangle = 0$ for all five simulations.}
\label{fig:cont_u_zeroline}
\end{figure}

In the statistical analysis to follow, the average of  a generic quantity $q$ is indicated with 
the angular brackets, $\langle q\rangle$, or with the capital letter, $Q$, according 
to convenience, while the fluctuation is indicated with the apex, 
$q'$. Details of the recirculation bubble for all the simulations are shown in 
figure~\ref{fig:cont_u_zeroline}, providing the contour plot of the mean streamwise 
velocity $\langle u_x \rangle$ normalised with the bulk velocity. The black solid line 
highlights the zero isolines to better appreciate the mean flow reversal. 

By progressively restricting the section, the first part of the bump makes the flow 
velocity increase consistently with the pressure decrease which is mechanically responsible 
for the acceleration. After the top of the bump the flow decelerates and a strong adverse 
pressure gradient occurs. This produces a backward flow near the bottom wall, giving rise 
to flow detachment. The bubble becomes smaller and more attached to 
the bump starting from the lower Reynolds number (A1) to the higher Reynolds number (A3).
The profiles recuperate positive average velocity at the bottom wall behind the 
bump at an earlier x-position for simulation A3 compared to simulation A2 or A1. 
Concerning the effect of the geometry, the bubble becomes larger starting from the 
streamlined bump in simulation A1 to the more bluff geometry in simulation C1. 
However, the mean position of the reattachment point in the streamwise direction is 
basically independent of the bump width, at $x=7$ for all three geometries at 
$\Rey = 2500$, while it clearly depends on the Reynolds number.

\begin{figure}
  \centerline{
  \includegraphics[width=0.32\textwidth]{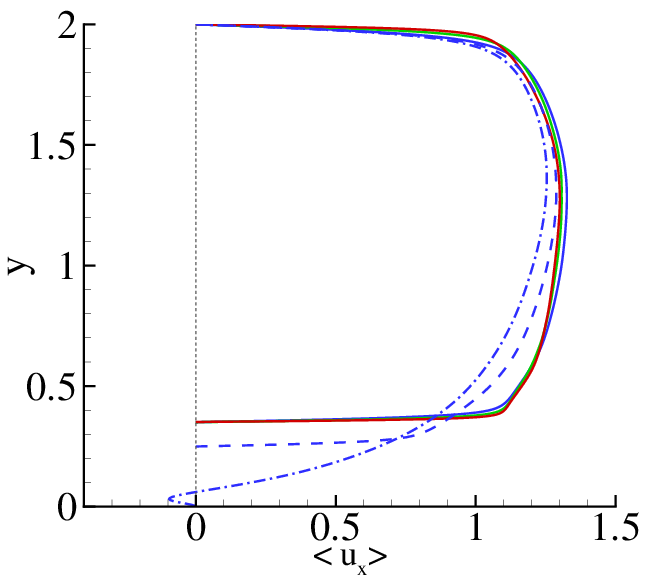}
    {\scriptsize \put(-125,102){\bf (a)}}
  \includegraphics[width=0.32\textwidth]{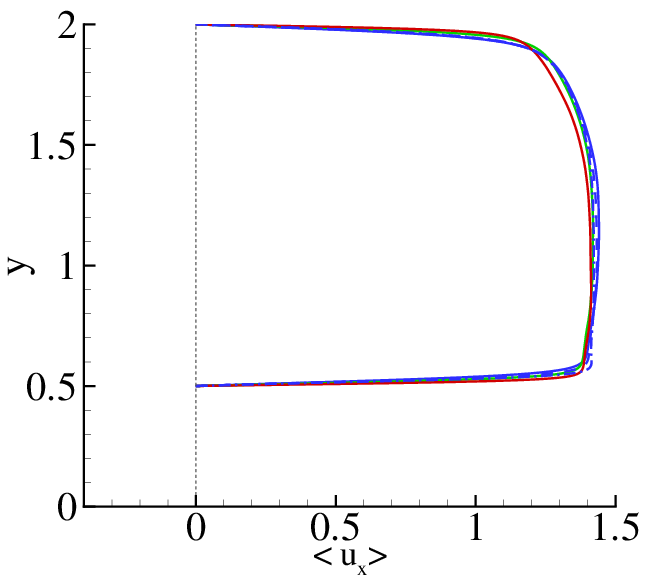}
    {\scriptsize \put(-125,102){\bf (b)}}
  \includegraphics[width=0.32\textwidth]{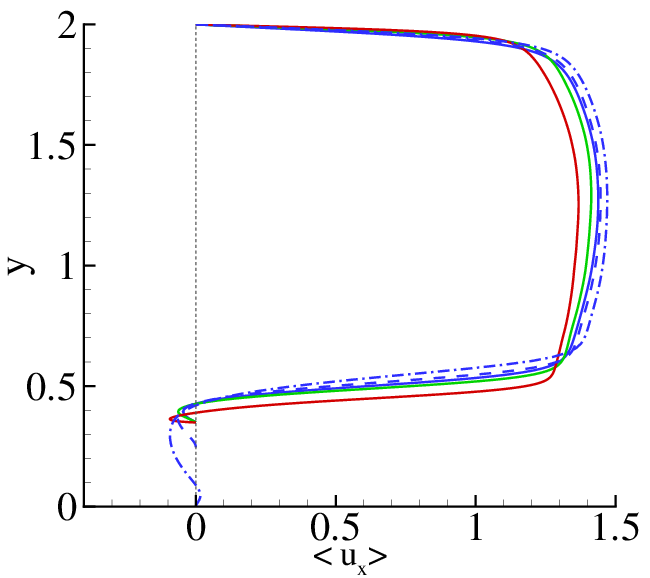}
    {\scriptsize \put(-125,102){\bf (c)}}}
      \centerline{
  \includegraphics[width=0.32\textwidth]{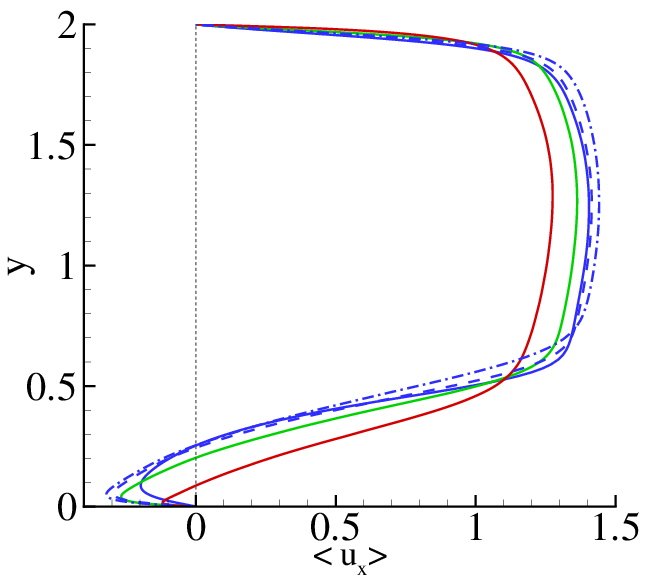}
    {\scriptsize \put(-125,102){\bf (d)}}
  \includegraphics[width=0.32\textwidth]{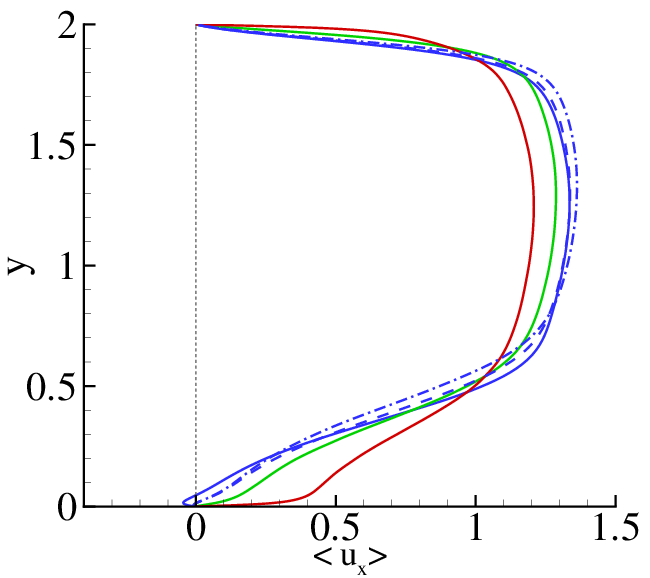}
    {\scriptsize \put(-125,102){\bf (e)}}
  \includegraphics[width=0.32\textwidth]{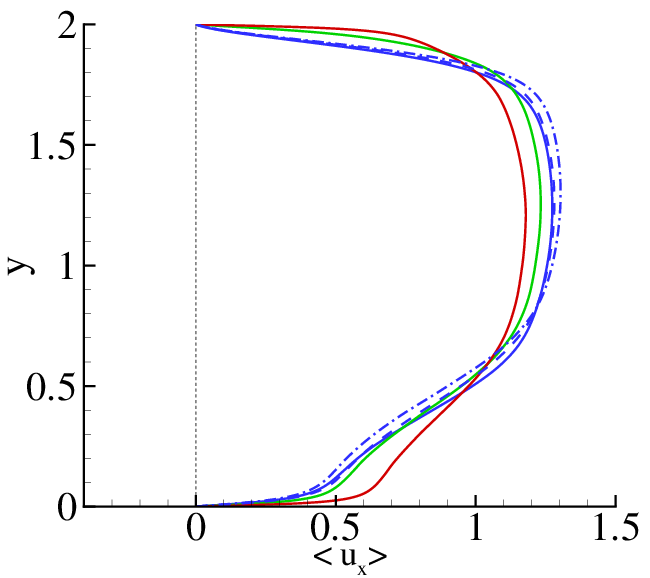}
    {\scriptsize \put(-125,102){\bf (f)}}}
  \caption{Mean axial velocity, $\langle u_x\rangle$, at six positions (a) to (f) 
   corresponding to the stations in figure~\ref{fig:stations}. Simulations A1, B1 and C1 
   are represented by solid, dashed and dashed-dotted blue lines respectively. Simulations 
   A2 and A3 are represented by solid green and red lines respectively.}
\label{fig:uprof_mean}
\end{figure}

\begin{figure}
  \centerline{
  \includegraphics[width=0.32\textwidth]{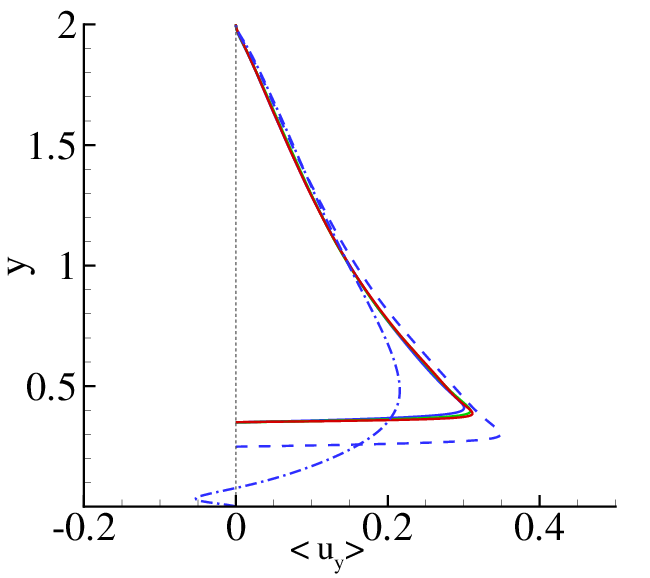}
    {\scriptsize \put(-125,102){\bf (a)}}
  \includegraphics[width=0.32\textwidth]{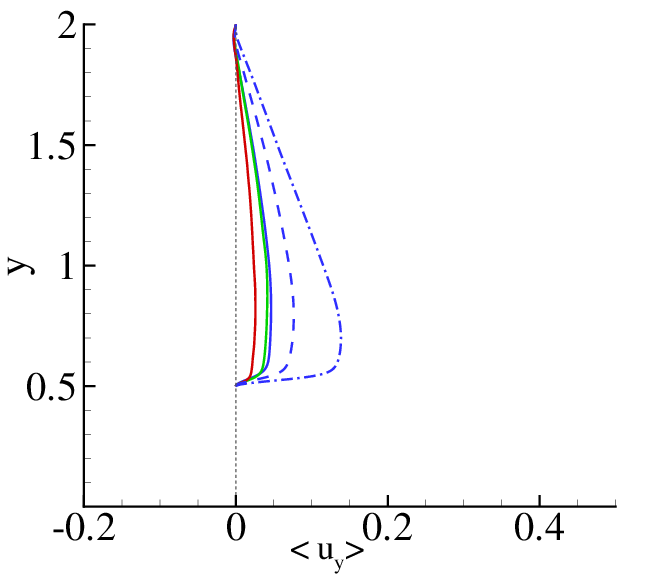}
    {\scriptsize \put(-125,102){\bf (b)}}
  \includegraphics[width=0.32\textwidth]{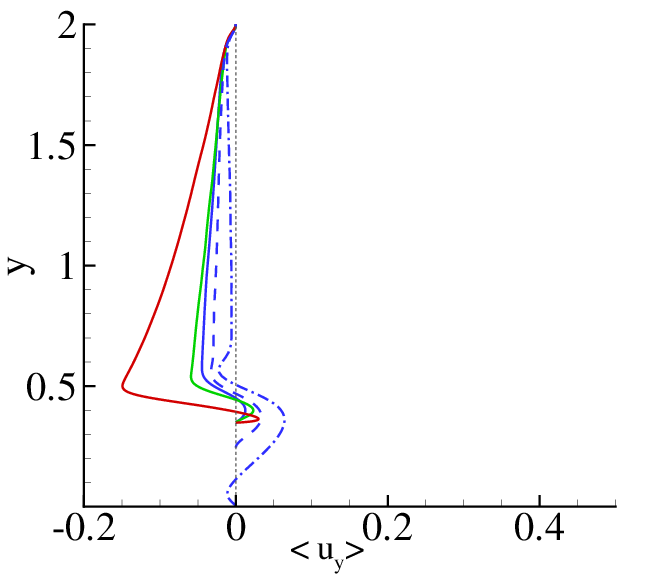}
    {\scriptsize \put(-125,102){\bf (c)}}}
      \centerline{
  \includegraphics[width=0.32\textwidth]{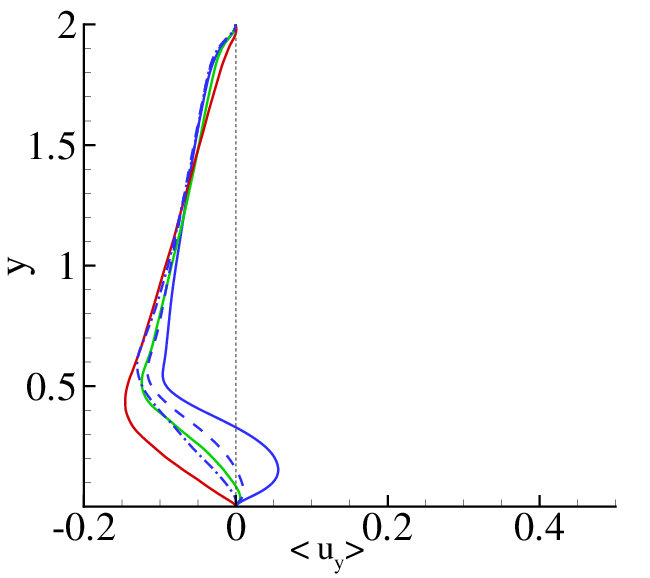}
    {\scriptsize \put(-125,102){\bf (d)}}
  \includegraphics[width=0.32\textwidth]{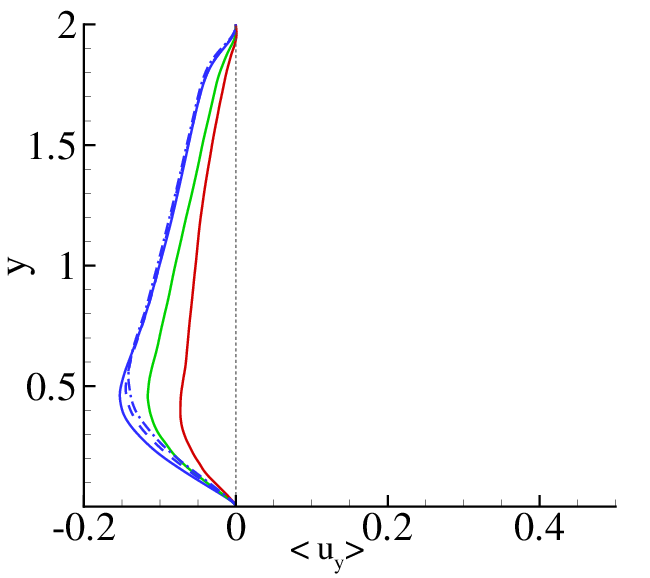}
    {\scriptsize \put(-125,102){\bf (e)}}
  \includegraphics[width=0.32\textwidth]{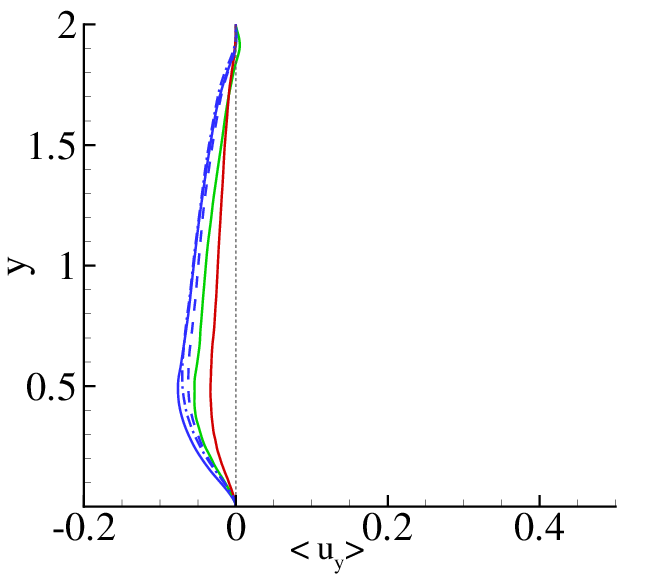}
    {\scriptsize \put(-125,102){\bf (f)}}}
  \caption{Mean wall-normal velocity, $\langle u_y\rangle$, at six positions (a) to (f) 
   corresponding to the stations in figure~\ref{fig:stations}. Simulations A1, B1 and C1 
   are represented by solid, dashed and dashed-dotted blue lines respectively. 
   Simulations A2 and A3 are represented by solid green and red lines respectively.}
\label{fig:vprof_mean}
\end{figure}

The above discussion is confirmed in detail by considering the mean velocity profiles 
extracted from figure~\ref{fig:cont_u_zeroline} and reported in figures~\ref{fig:uprof_mean}
and~\ref{fig:vprof_mean} at the downstream positions shown in figure~\ref{fig:stations}. 
Simulations A1, B1 and C1 are represented by solid, dashed and dashed-dotted blue lines 
respectively whilst simulations A2 and A3 are represented by solid green and red lines 
respectively. Figure~\ref{fig:uprof_mean} shows mean streamwise velocity profiles for all 
five simulations. For the station shown in panel (a), the profiles are almost independent of 
the Reynolds number (solid lines) while they depend strongly on the geometry (broken lines), 
suggesting that the mean flow has already achieved a sort of asymptotic state. The profile 
for case C1 extends down to the bottom wall with a slightly negative velocity, 
indicating a small recirculating region ahead of the bump. At the tip of the bump, panel (b), 
all the profiles essentially exhibit the same behaviour. The recirculation is already well 
developed at the station of panel (c) for the bluffest bulge, case C1.  Further downstream, 
panel (d), the wall-normal extension of the backward flow are well evident for all cases
except for the highest Reynolds number (case A3), where the recirculation is more attached to 
the wall and extends less streamwise. Since the extension of the bubble is larger for the 
lower Reynolds number cases, it is still present at the station corresponding to panel (e). 
In contrast, reattachment already occurred for the higher Reynolds number (cases A2-A3). 
Further downstream, panel (f), the flow is attached for all conditions. 

Figure~\ref{fig:vprof_mean} shows the mean wall-normal velocity profiles. 
Wall-normal velocities are particularly intense as the flow invests the bump, panel (a). 
At the tip of the bump, panel (b), the positive (away from the bottom wall) wall-normal 
velocity peak is higher for the less streamlined  bump (C1), indicating that the flow 
is strongly converging towards the opposite wall, leading to a contraction of the 
effective section ({\sl vena contracta}). The wall-normal velocity is progressively reduced 
downstream, to eventually become negative. At intermediate stations, see e.g. panel (c), 
the wall-normal velocity is negative in the outer flow, indicating the trend toward 
reattachment to the lower wall. Approaching the wall, $\langle u_y \rangle$ becomes zero at 
the edge of the recirculation bubble. Inside the bubble $\langle u_y \rangle$ is positive, 
indicating that the profile is traversing the fore part of the bubble, 
recirculating clockwise. Moving further downstream, the external flow still moves 
toward the lower wall, but now the aft part of the bubble is reached, implying a 
negative wall-normal velocity also inside the bubble. Finally, the reattachment point 
is reached and the wall-normal average velocity tends to vanish in the entire channel 
section, starting to recover nearly parallel-flow conditions expected far away from the bump.
The shorter bubble length leads to a more abrupt reattachment, as seen by the large 
negative wall-normal average velocity in the red plot of panel (c). The bluffest 
configuration induces an evident secondary recirculation bubble just below the primary 
one where the bump ends, see also figure~\ref{fig:cont_u_zeroline}. 
Note that in this figure a small bubble is also apparent just ahead of the bump.

\begin{figure}
  \centerline{
  \includegraphics[width=0.32\textwidth]{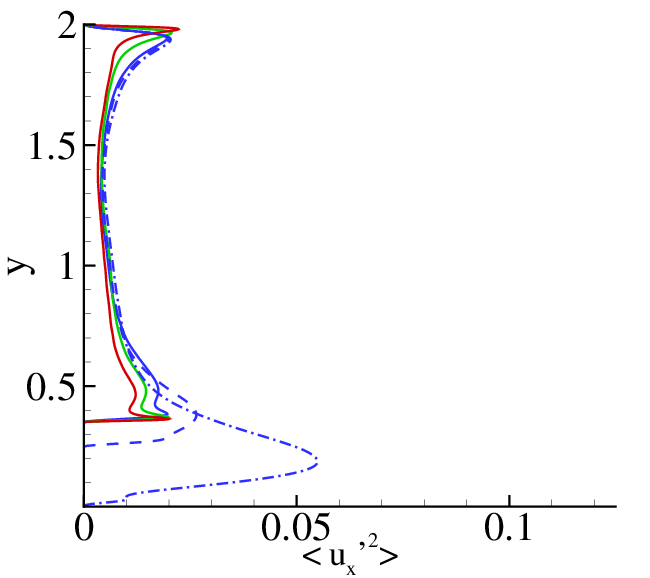}
    {\scriptsize \put(-125,102){\bf (a)}}
  \includegraphics[width=0.32\textwidth]{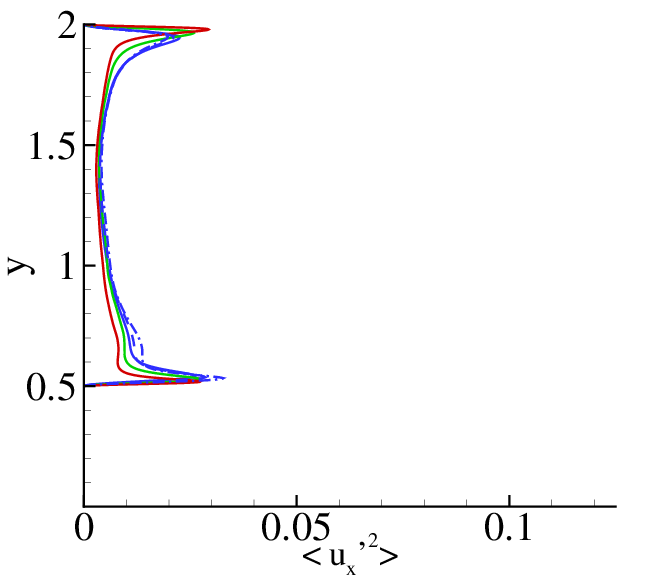}
    {\scriptsize \put(-125,102){\bf (b)}}
  \includegraphics[width=0.32\textwidth]{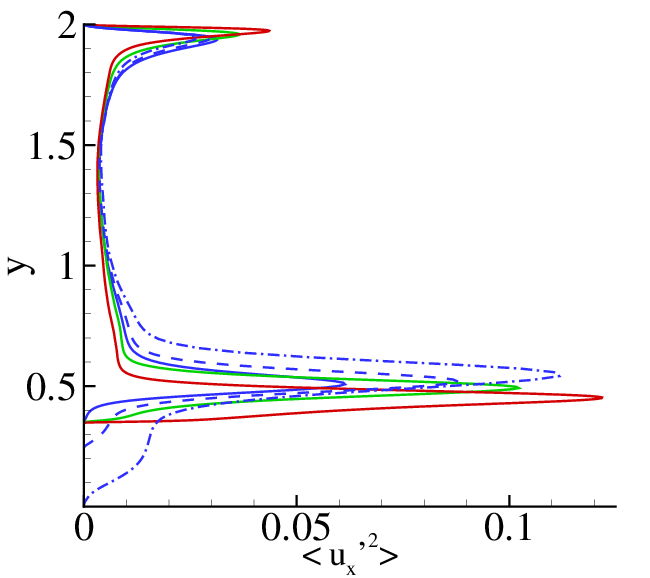}
    {\scriptsize \put(-125,102){\bf (c)}}}
      \centerline{
  \includegraphics[width=0.32\textwidth]{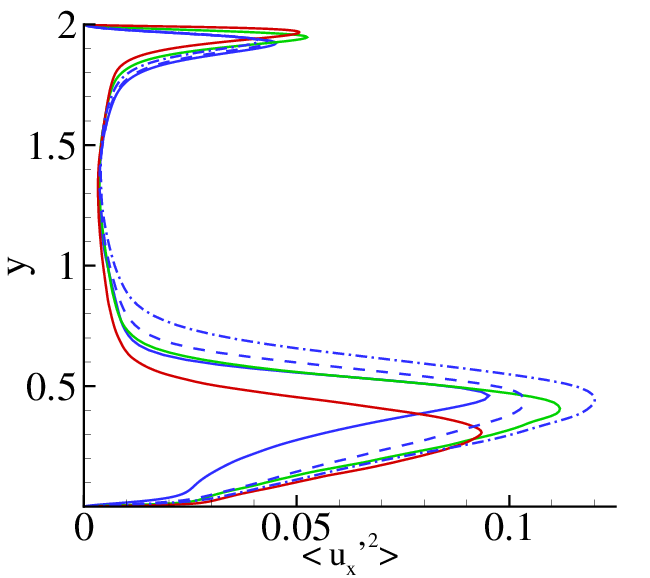}
    {\scriptsize \put(-125,102){\bf (d)}}
  \includegraphics[width=0.32\textwidth]{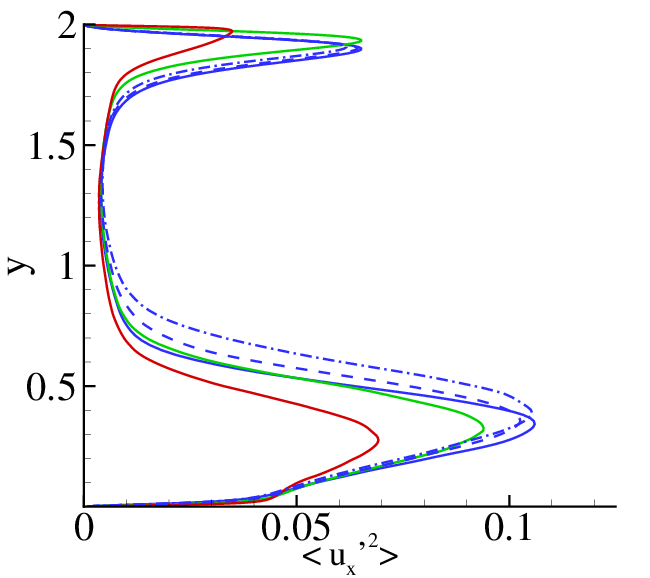}
    {\scriptsize \put(-125,102){\bf (e)}}
  \includegraphics[width=0.32\textwidth]{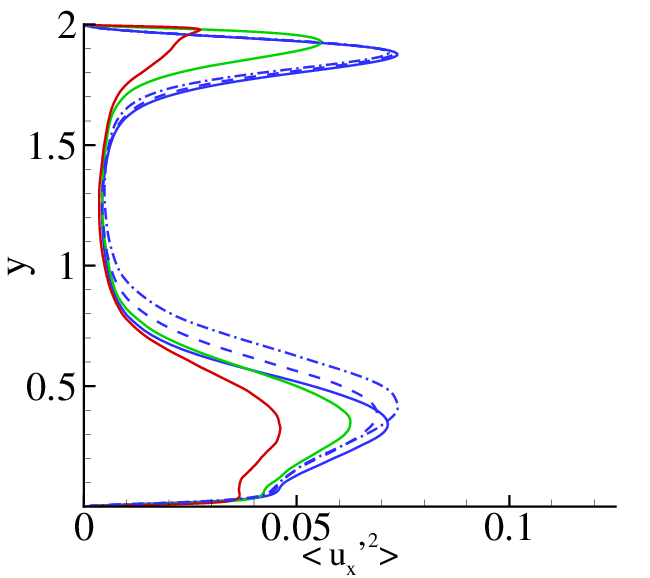}
    {\scriptsize \put(-125,102){\bf (f)}}}
  \caption{Mean streamwise velocity fluctuations, $\langle u_x'^2\rangle$, at six 
   positions (a) to (f) corresponding to the stations in figure~\ref{fig:stations}. 
   Simulations A1, B1 and C1 are represented by solid, dashed and dashed-dotted 
   blue lines respectively. Simulations A2 and A3 are represented by solid green 
   and red lines respectively.}
\label{fig:uxF2prof_mean}
\end{figure}
\begin{figure}
  \centerline{
  \includegraphics[width=0.32\textwidth]{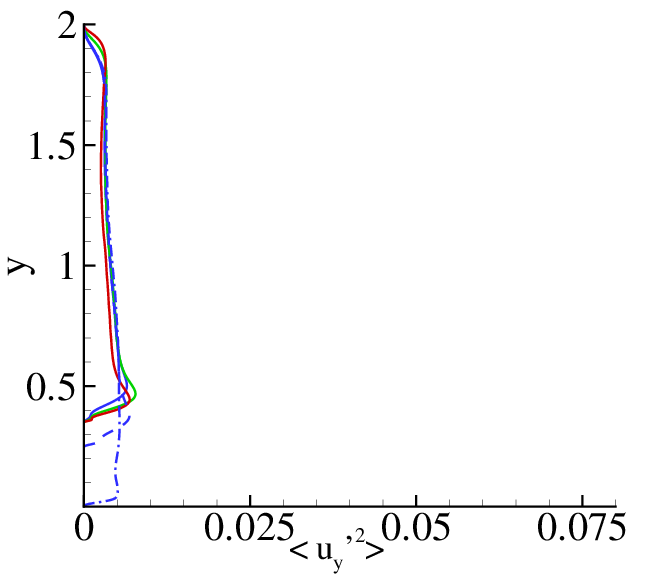}
    {\scriptsize \put(-125,102){\bf (a)}}             
  \includegraphics[width=0.32\textwidth]{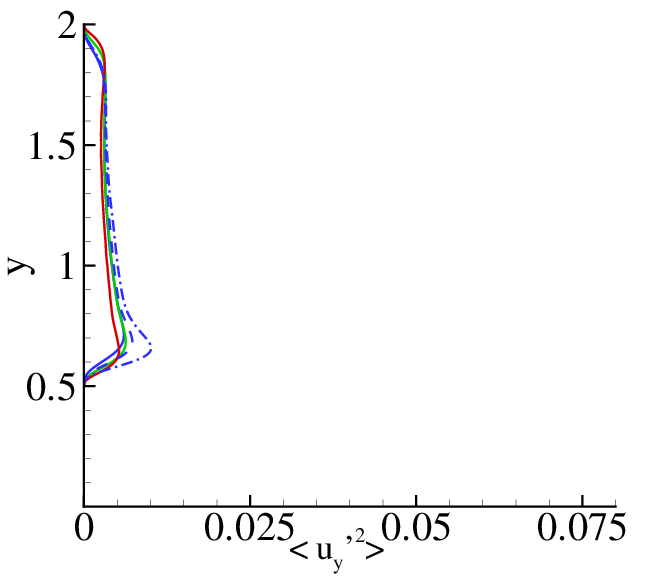}
    {\scriptsize \put(-125,102){\bf (b)}}            
  \includegraphics[width=0.32\textwidth]{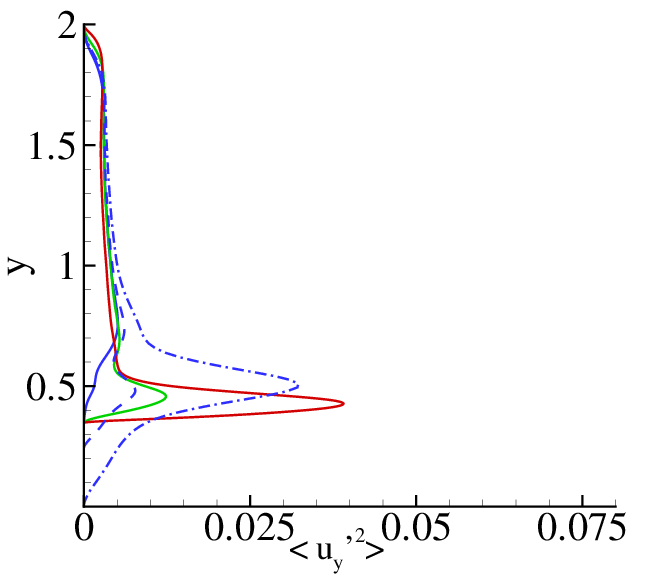}
    {\scriptsize \put(-125,102){\bf (c)}}}          
      \centerline{                                 
  \includegraphics[width=0.32\textwidth]{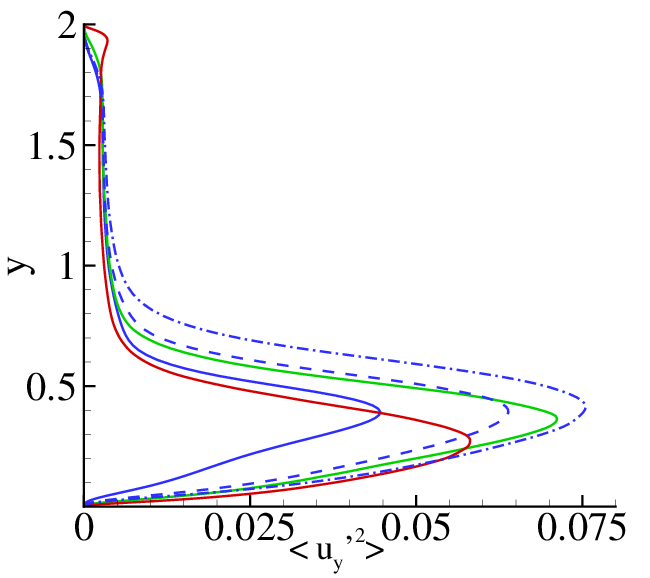}
    {\scriptsize \put(-125,102){\bf (d)}}         
  \includegraphics[width=0.32\textwidth]{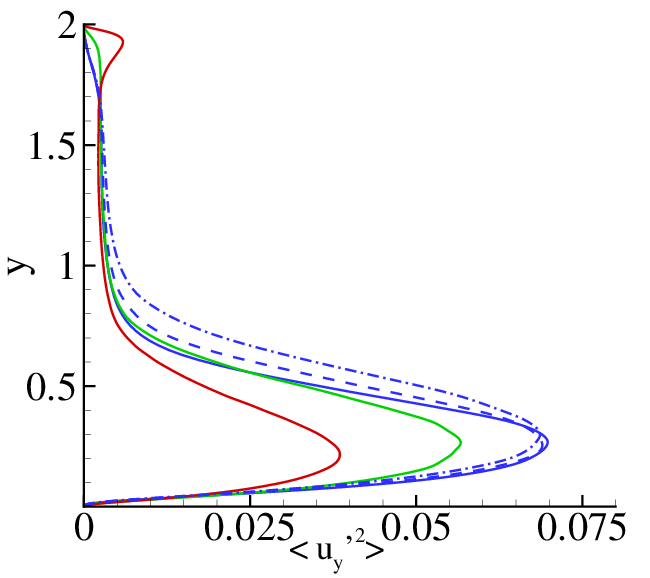}
    {\scriptsize \put(-125,102){\bf (e)}}        
  \includegraphics[width=0.32\textwidth]{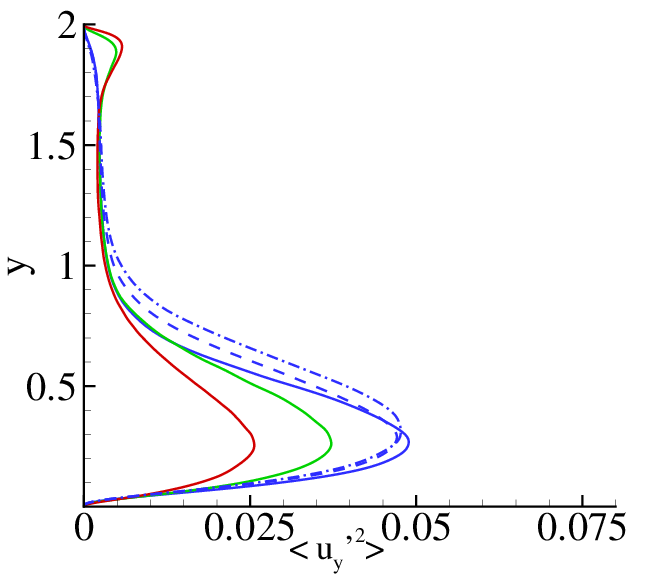}
    {\scriptsize \put(-125,102){\bf (f)}}}
  \caption{Mean wall-normal velocity fluctuations, $\langle u_y'^2\rangle$, at six positions (a) to (f) corresponding to the stations in 
figure~\ref{fig:stations}. 
Simulations A1, B1 and C1 are represented by solid, dashed and dashed-dotted blue lines respectively. 
Simulations A2 and A3 are represented by solid green and red lines respectively.}
\label{fig:uyF2prof_mean}
\end{figure}
\begin{figure}
  \centerline{
  \includegraphics[width=0.32\textwidth]{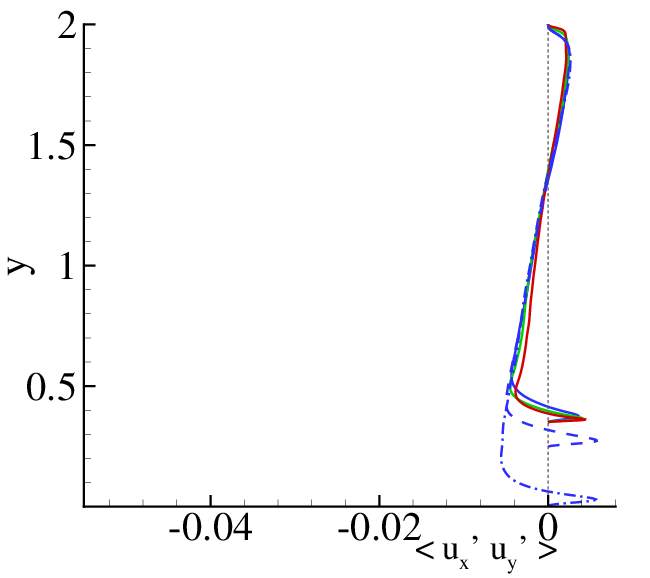}
    {\scriptsize \put(-125,102){\bf (a)}}             
  \includegraphics[width=0.32\textwidth]{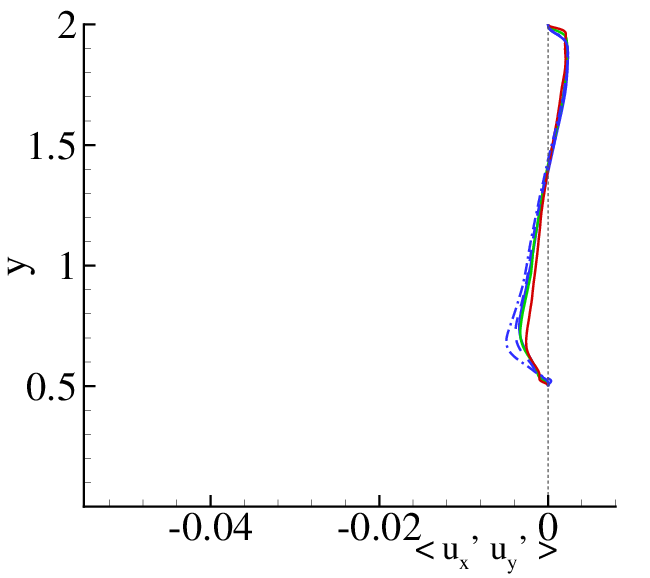}
    {\scriptsize \put(-125,102){\bf (b)}}            
  \includegraphics[width=0.32\textwidth]{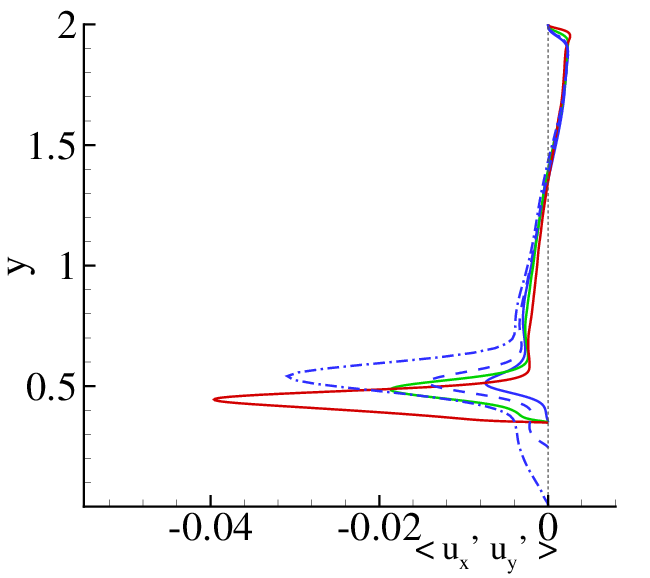}
    {\scriptsize \put(-125,102){\bf (c)}}}          
      \centerline{                                 
  \includegraphics[width=0.32\textwidth]{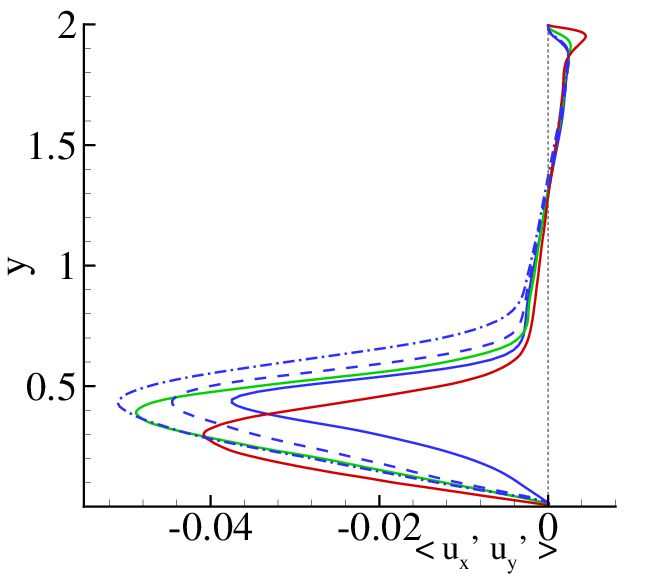}
    {\scriptsize \put(-125,102){\bf (d)}}             
  \includegraphics[width=0.32\textwidth]{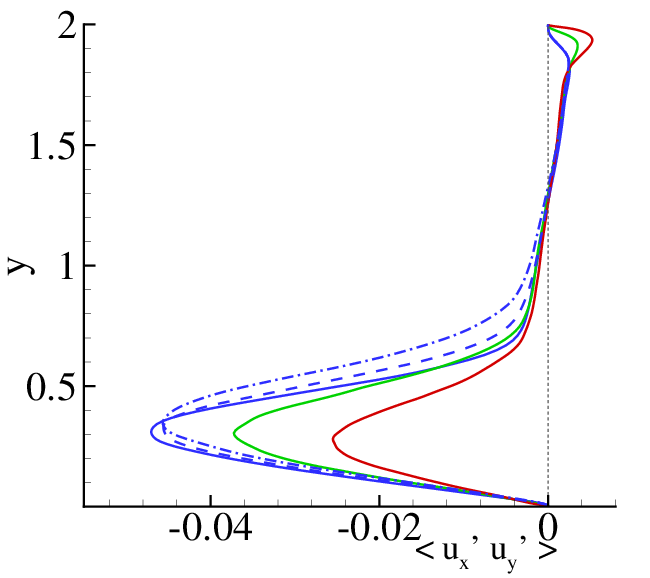}
    {\scriptsize \put(-125,102){\bf (e)}}            
  \includegraphics[width=0.32\textwidth]{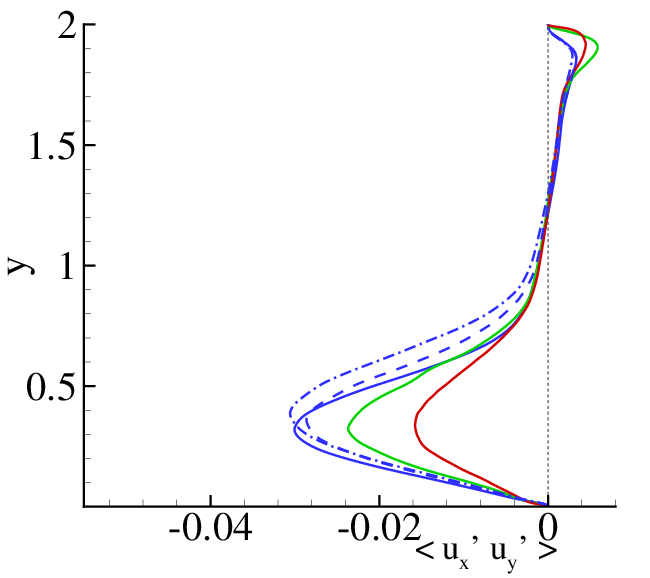}
    {\scriptsize \put(-125,102){\bf (f)}}}
  \caption{Reynolds stress, $\langle u_x'u_y'\rangle$, at six positions (a) to (f) corresponding to the stations in 
figure~\ref{fig:stations}. 
Simulations A1, B1 and C1 are represented by solid, dashed and dashed-dotted blue lines respectively. 
Simulations A2 and A3 are represented by solid green and red lines respectively.}
\label{fig:uxFuyFprof_mean}
\end{figure}

The next figures present the mean profiles for the fluctuating quantities, i.e.
$\langle u_x'^2 \rangle$, $\langle u_y'^2 \rangle$ and $\langle u_x' u_y' \rangle$,
at the same stations addressed above for the mean velocity profiles.
Figure~\ref{fig:uxF2prof_mean} shows mean streamwise velocity fluctuation profiles 
$\langle u_x'^2 \rangle$ for all five simulations. These are particularly strong close to the walls or inside 
the shear layer above the recirculating region, for all cases. The fluctuations are 
maximum in the high Reynolds number case (A3) and in the most bluff case (C1). In the 
former, the high velocity fluctuations are due to the higher Reynolds number, whilst 
in the latter, the fluctuations are fed by the strong flow separation induced by the 
bump shape. In the lower Reynolds number simulations, the regions with higher
velocity fluctuations are larger, corresponding to a larger shear layer.
The presence of the bump also results in an increase in fluctuations at the top wall, 
particularly evident for the lower Reynolds number simulations. 
The change in geometry also affects the maxima reached by the fluctuations. 
In particular, due to the increased size of the recirculating region, simulation C1 
exhibits peak values which are comparable to simulation A3 in the shear layer, 
see panel (c). 
In correspondence of the recirculating region,
the peaks are higher for case C1 with respect to the cases at the same Reynolds number, B1 and A1.
The profiles for the wall-normal velocity fluctuations $\langle u_y'^2 \rangle$ 
and the Reynolds stress, $\langle u_x'u_y'\rangle$, follow a similar trend.
The profiles are shown in figure~\ref{fig:uyF2prof_mean} and 
figure~\ref{fig:uxFuyFprof_mean} for all five simulations. The positive values of 
Reynolds stress confirm the presence of the small recirculation bubble 
ahead of the bump, panel (a) in figure~\ref{fig:uxFuyFprof_mean}.

\begin{figure}
\centering
\includegraphics[width=0.5\textwidth]{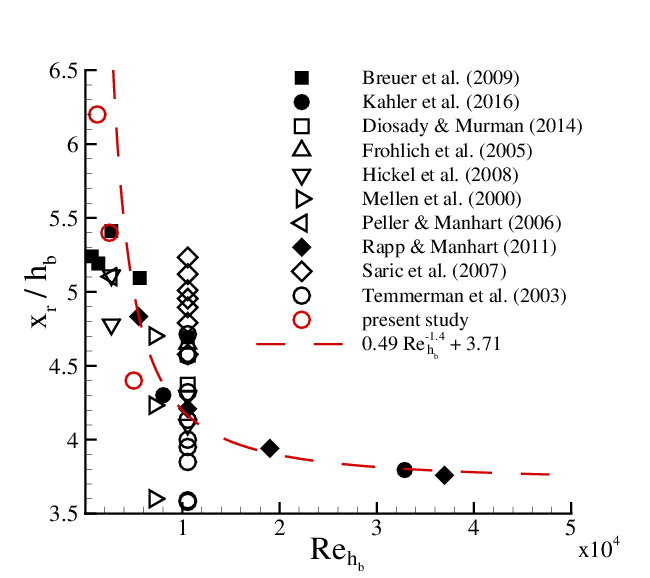}
\caption{\label{fig:comp_exp} Dependence of the reattachment point normalised with the height of 
the obstacle $x_r/h_b$, on the Reynolds number based on $h_b$, $\Rey_{h_b}=\Rey\, h_b$. 
$h_b$ is the bump height normalised with the nominal channel half-height, $h_0$. 
Closed black symbols refer to experimental measurements, open black symbols to numerical simulations 
(mainly LES). Red dashed line is the exponential fit of the experimental data~\cite{Kahler_2016}. 
Open red symbols are extracted from the present DNSs. \cite{breuer2009flow}, \cite{Kahler_2016},
\cite{diosady2014dns}, \cite{frohlich2005highly}, \cite{hickel2008implicit}, \cite{mellen2000LES}, \cite{peller2006turbulent},
\cite{Rapp_2011}, \cite{saric_2007}, \cite{temmerman2003investigation}}
\end{figure}
The position of the reattachment point clearly does not depend on the dimension of the 
obstacle in the streamwise direction, but on the Reynolds number. These observations are 
in agreement with both numerical and experimental data already available in 
the literature. Figure~\ref{fig:comp_exp} shows the reattachment point normalised with 
the height of the bump, $x_r/h_b$ as a function of the Reynolds number based on the 
bulk velocity and on the height of the bump, $\Rey_{h_b}=\Rey\, h_b$ (note that $h_b$ is
the dimensionless bump height). Data have been collected in~\cite{Kahler_2016} from 
several experiments (closed symbols) and numerical simulations, mainly LES, 
(open symbols), whilst the open red circles are the present simulations.

Note that the geometries in these experiments and simulations only qualitatively corresponds to ours. We therefore 
focus the comparison on the re-attachment point whose position is mainly controlled by bump height 
and Reynolds number, quite independently of the detailed geometry.
The red dashed line in the figure is a power-law fit, based on all experimental results, 
which asymptotically reaches $3.71\,x/h_b$ for large Reynolds numbers. The reattachment 
location scales with $\Rey_{h_b}$ to the power $-1.4$. The reattachment position 
moves further upstream with increasing Reynolds number due to the stronger turbulent 
mixing, i.e. due to a higher turbulent momentum transfer towards the wall. The present 
DNS data agree well with the experimental results of \cite{Kahler_2016} and of \cite{Rapp_2011}.
Overall, this compilation of data shows a significant scatter of data obtained by turbulence modeling 
and a certain inability of the models to capture the bubble  reattachment position, see also 
the discussion in~\cite{Kahler_2016} for more details. Concerning the present DNS, the slight 
differences with the experiments can be attributed to details of the turbulence investing the bump 
and the confinement effect of the upper wall.

\subsection{Pressure, drag and friction coefficients}

% Cd
The instantaneous pressure is decomposed as the sum of a contribution linearly 
decreasing in the streamwise direction, which is associated to the instantaneous 
pressure drop $\Delta p(t)$ across the channel, plus a departure $\tilde p$ from the linear 
law,
$$
p(\vx,t)  = -\frac{\Delta p}{L_x} x + {\tilde p}(\vx,t) \ .
$$
In the present simulations $\Delta p(t)$ does not significantly 
fluctuate in time, oscillating within 1\% at most, although its value is in principle continuously adjusted to 
keep the flow rate rigorously constant. The drag coefficient in terms of the present dimensionless 
variables is
$$
C_d =4\frac{\Delta P}{L_x}=-\frac{2}{L_x}\int_{walls}\langle {\bf t}\rangle \cdot {\bf e}_x  dl\, ,
$$
where $\Delta P = \langle \Delta p\rangle$ is the average pressure drop,  
$\bf t$ is the (dimensionless) traction at the wall (pressure plus shear force) and 
${\bf e}_x$ is the unit vector in the streamwise direction. Wherever needed (see e.g. \S\ref{sec:TKE})
the fluctuation of the pressure drop will be denoted by $\Delta p'(t)$.

The drag increases moving from the most streamlined (A1) to the most bluff profile (C1), 
see table~\ref{TABLE}. On the other hand, the drag coefficient decreases with the increase in 
Reynolds number, from simulation A1, A2 to A3.  For purpose of comparison, the drag in a planar 
channel at the same flow rate is $C_d^{channel} = 4 \left(\Rey_\tau^0/\Rey_b\right)^2$. 
The drag coefficient can be decomposed in form and friction components, namely 
$$
C_d^{form} =  \frac{2}{L_x} {\bf e}_x \cdot \int_{walls}  P\, {\bf n}\,  dl\, , \qquad 
C_d^{friction} =  -\frac{2}{L_x} {\bf e}_x \cdot \int_{walls} \mu \, \frac{\partial {\bf U}}{\partial n}  dl\, ,
$$
where ${\bf n}$ is the unit normal exiting the fluid domain. The form drag coefficient increases by 
$50 \%$ going from the most streamlined to most bluff shape.
The observed decrease of the form drag coefficient with increasing Reynolds number is more than compensated by the larger 
velocity for flows in the same geometry implying, as obvious, the increase of the corresponding 
contribution to the resistance, $D^{form} \propto U_b^2 C_d^{form}$. In the present geometry, the friction component is 
dominated by the straight part of the channel and its value is not significantly different from the one expected in a planar channel,
see the small difference $C_d^{friction}-C_d^{channel}$ in table~\ref{TABLE}. This confirms that most of the bump-induced 
drag should be interpreted as form drag.

\begin{table}
\centering
\begin{tabular}{cccccccccc}
\hline
Case && $C_d$ && $C_d^{form}$ && $C_d^{channel}$  && $C_d^{friction}-C_d^{channel}$   \\
\hline
$A1$ && $3.07 \cdot 10^{-2}$&&$1.06 \cdot 10^{-2}$ && $1.80 \cdot 10^{-2}$ && $0.21 \cdot 10^{-2}$  \\
$B1$ && $3.26 \cdot 10^{-2}$&&$1.28 \cdot 10^{-2}$ && $1.80 \cdot 10^{-2}$ && $0.18 \cdot 10^{-2}$  \\
$C1$ && $3.49 \cdot 10^{-2}$&&$1.55 \cdot 10^{-2}$ && $1.80 \cdot 10^{-2}$ && $0.14 \cdot 10^{-2}$  \\
$A2$ && $2.68 \cdot 10^{-2}$&&$1.05 \cdot 10^{-2}$ && $1.44 \cdot 10^{-2}$ && $0.19 \cdot 10^{-2}$  \\
$A3$ && $2.05 \cdot 10^{-2}$&&$7.07 \cdot 10^{-3}$ && $1.17 \cdot 10^{-2}$ && $0.17 \cdot 10^{-2}$  
\end{tabular}
\caption{ \label{TABLE} Drag coefficient decomposed into form and friction contributions 
and comparison against an equivalent planar channel. See text for definitions.
}
\end{table}

% Pressure 2D Plots
%
\begin{figure}
  \centerline{\includegraphics[width=0.7\linewidth]{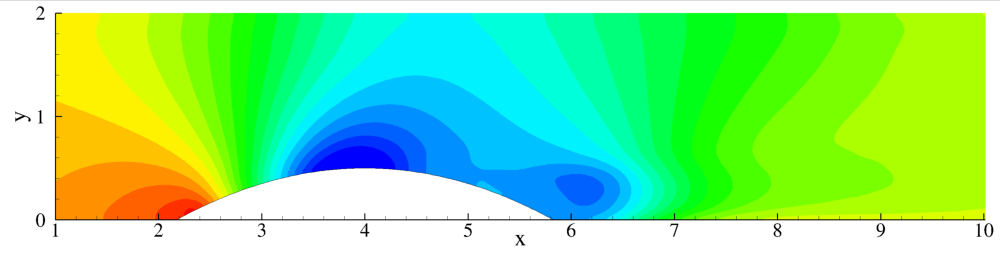}
  {\scriptsize \put(-300,63){\bf (A1)}}}
  \centerline{\includegraphics[width=0.7\linewidth]{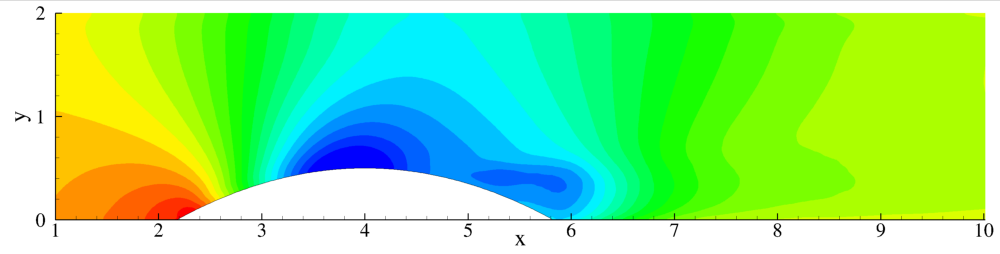}
  {\scriptsize \put(-300,63){\bf (A2)}}}
  \centerline{\includegraphics[width=0.7\linewidth]{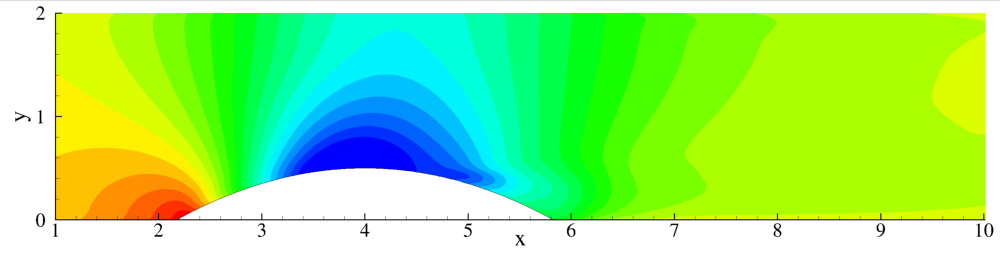}
  {\scriptsize \put(-300,63){\bf (A3)}}}
  \centerline{\includegraphics[width=0.7\linewidth]{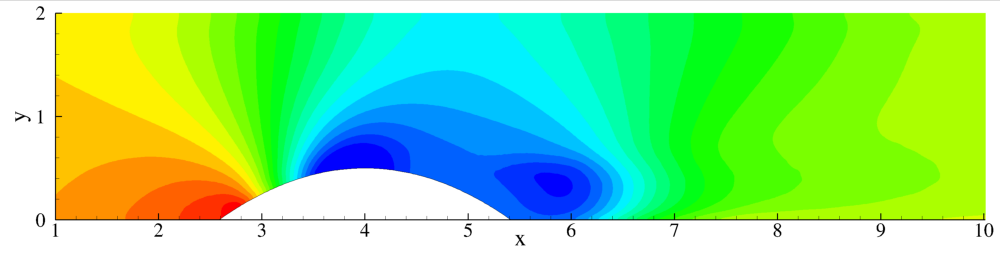}
  {\scriptsize \put(-300,63){\bf (B1)}}}
  \centerline{\includegraphics[width=0.7\linewidth]{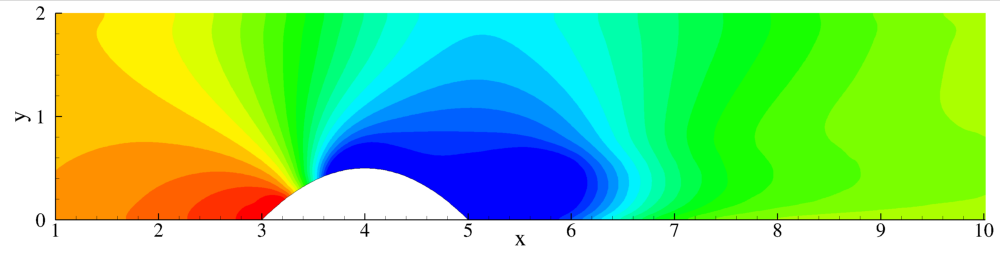}
  {\scriptsize \put(-300,63){\bf (C1)}}}
  \centerline{\includegraphics[width=0.3\linewidth]{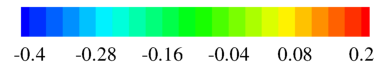}       }
  \caption{Average pressure $\langle p \rangle$ contour plots for all five simulations.}
\label{fig:cont_p}
\end{figure}

Given its relevance in determining the drag, the  mean pressure field is shown
in figure~\ref{fig:cont_p} for all simulations. Roughly, the qualitative behaviour is similar 
for all cases. A high pressure region occurs just before 
the bump where a stagnation point occurs. Further downstream the 
pressure decreases reaching its minimum in the high velocity region,
at the top of the bump. The pressure field behind the bump is 
strongly influenced by the shape and dimension of the recirculation bubble. 
When the separation region is significant, cases A1, B1 and C1, a second 
pressure minimum develops inside the recirculation bubble. 
The high Reynolds number simulations where performed on the most streamlined geometry, 
which produces a smaller separation compared to the other geometries.
With increasing Reynolds number, cases A2 and A3, the flow more easily faces the 
adverse pressure gradient due to  enhanced turbulent mixing, resulting in delayed separation.  
As a consequence, pressure recovery behind the bump is more effective.
% Cp
%
\begin{figure}
  \centerline{\includegraphics[width=0.9\linewidth]{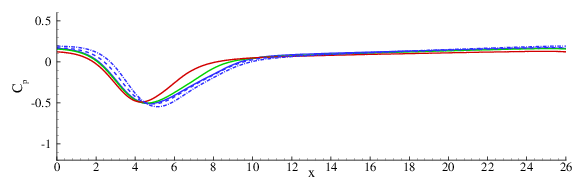}}
  \centerline{\includegraphics[width=0.9\linewidth]{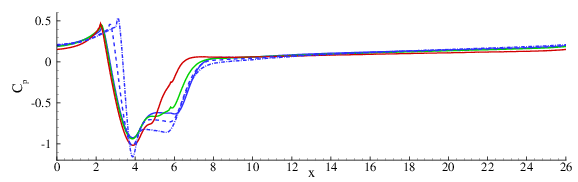}}
  \caption{Modified pressure coefficient at the top and bottom walls of the domain,
           top and bottom panels respectively. Simulations A1, B1 and C1 are 
           represented by solid, dashed and dashed-dotted blue lines respectively. 
           Simulations A2 and A3 are represented by solid green and red lines 
           respectively.}
\label{fig:Cp}
\end{figure}

Given the average pressure drop along the channel, the effect of the bump on the wall 
pressure is better addressed in terms of a departure-pressure coefficient 
$$C_p^{t/b}(x) = \left.2\, \langle {\tilde p} \rangle\right|_{y=2/y=0}\, ,$$
where the superscripts `$t$' and `$b$' refer to the top and bottom wall respectively.  
Adding the linear term $-2 x \Delta P/L_x$ recovers the standard definition of $C_p$.  
Figure~\ref{fig:Cp} shows the departure-pressure at both walls.  The effect of the bump extends to the 
opposite wall, with $C_p^t$ presenting a trough just after the bump tip (at $x=4$) and recovering
downstream. The trough is smaller and shifted closer to the bump tip with increasing Reynolds number.
At constant Reynolds number, the trough is more pronounced and farther away from the tip 
when the geometry is bluffer, i.e. simulations B1 and C1. 
At the bottom wall, as the flow reaches the bump, $C_p^b$ initially increases producing  a small 
recirculation just ahead of the bump. The pressure then decreases to its minimum 
slightly ahead of the tip. These trends become stronger for the bluffer geometries.
Immediately downstream of the tip, the pressure abruptly rises reaching the separation point.
In the recirculation bubble the pressure remains almost constant, with the extension of the plateaux
becoming smaller at increasing Reynolds number (red line). The extension of the plateaux
is almost independent of the geometry, where $C_p^b$ decreases for bluffer bumps.

% Cf
%
\begin{figure}
  \centerline{\includegraphics[width=0.9\linewidth]{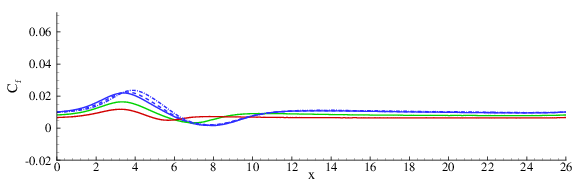}}
  \centerline{\includegraphics[width=0.9\linewidth]{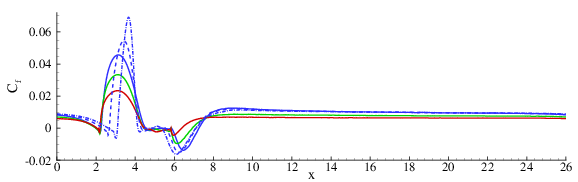}}
  \caption{Mean skin friction coefficient along $x$ at the top and bottom walls of the domain in the top and bottom panels respectively.
  Simulations A1, B1 and C1 are represented by solid, dashed and dashed-dotted blue lines respectively. 
  Simulations A2 and A3 are represented by solid green and red lines respectively.}
\label{fig:Cf}
\end{figure}
Figure~\ref{fig:Cf} shows the mean skin friction coefficient $C_f = 2 \, \tau_{w}$. At the top wall, 
$C_f$ is always positive, showing that the adverse pressure gradient (see figure~\ref{fig:Cp}) is 
too mild to induce average flow separation on the wall opposite to the bump. The maximum 
and minimum $C_f$ occur just before the bump tip and at the end of the bubble, respectively. 
At the bottom wall, the skin friction coefficient ahead of the bump 
becomes slightly negative due to the small recirculation bubble and reaches a positive peak as 
the flow approaches the tip of the bump. For the different Reynolds numbers, the 
position of the peak coincides but the maximum reduces with increasing Reynolds 
number. The peak is shifted towards the tip with increasing bluffness of geometry. 
A skin friction plateaux is observed behind the tip, along the bump where the cross-stream section 
of the channel increases. At the bubble, consistently with the backward flow at the wall, the skin friction 
is negative, with increasing absolute value for bluffer geometries and lower Reynolds numbers. 

%____________________________________________________________________________________________%
\subsection{Mean kinetic energy and turbulent kinetic energy budgets}

The Reynolds decomposition entails the splitting of the total kinetic energy in 
two parts, $K=K_M+k_T$, where
$K_M=1/2\, \langle u_i\rangle \langle u_i\rangle$ is the kinetic energy of the mean flow and 
$k_T=1/2 \langle u'_i u'_i \rangle$ is the turbulent kinetic energy. In literature, little  attention 
is typically paid to the kinetic energy of the mean field and most interest is focused on the 
turbulent contribution. This is motivated by the usually simple flow configuration where the 
mean balance equation for $K_M$ is trivial. In the present case both mean and turbulent 
kinetic energy need to be dealt with explicitly. The reason is that the bump breaks the 
streamwise homogeneity and induces strong mean wall-normal velocities. This gives rise 
to non-trivial  mean and turbulent spatial energy fluxes,  dissipation, and turbulent kinetic 
energy production.

\subsubsection{Mean kinetic energy}

The (stationary) mean flow kinetic energy equation reads
\begin{equation}
\pd{{\Phi_M}_j}{x_j} = -\varepsilon_M - {\Pi} + \frac{\Delta P}{L_x} U_x\, ,
\label{eq:MKE_all}
\end{equation}
where $\varepsilon_M = 1/\Rey (\partial U_i/\partial x_j) (\partial U_i/\partial x_j)$ 
is the mean flow energy dissipation rate per unit volume and 
${\Pi}= - \langle u'_i u'_j\rangle {\partial U_i }/{ \partial x_j}$ is the turbulent 
kinetic energy production. ${ U_x \Delta P}/{L_x}$ is the external average power input. 
The spatial flux,
\begin{equation}
{\Phi_M}_j = U_j K_M + {U_j \langle {\tilde p}\rangle} -\frac{1}{\Rey}\pd{K_M}{x_j} + U_i \langle u'_i u'_j\rangle \, ,
\label{eq:MKE_flux}
\end{equation}
redistributes energy across the flow, overall providing zero net contribution to the power. 

\begin{figure}
  \centerline{\includegraphics[width=0.7\linewidth]{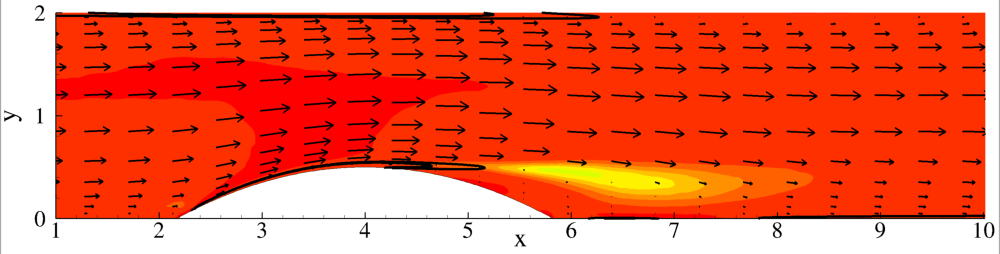}
  {\scriptsize \put(-300,63){\bf (A1)}}}
  \centerline{\includegraphics[width=0.7\linewidth]{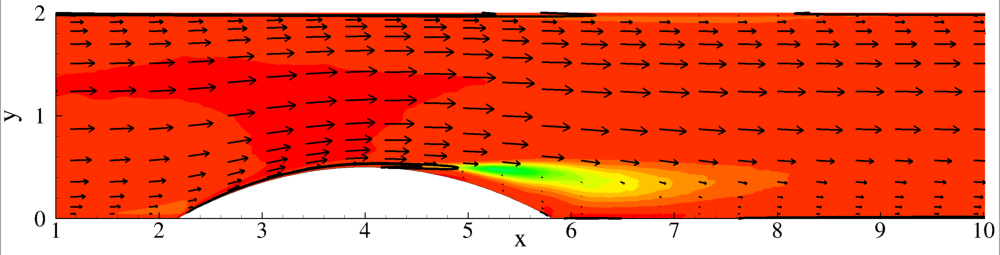}
  {\scriptsize \put(-300,63){\bf (A2)}}}
  \centerline{\includegraphics[width=0.7\linewidth]{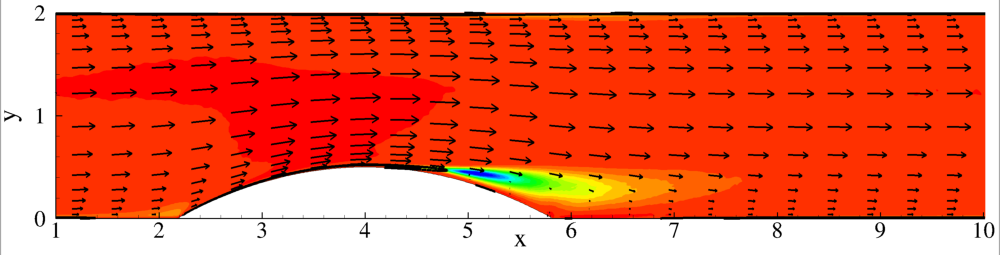}
  {\scriptsize \put(-300,63){\bf (A3)}}}
  \centerline{\includegraphics[width=0.7\linewidth]{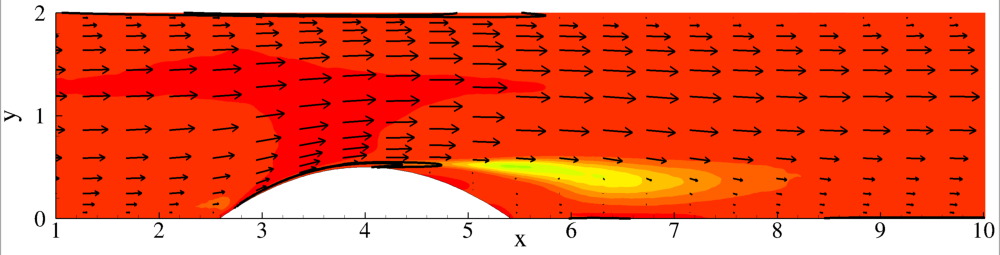}
  {\scriptsize \put(-300,63){\bf (B1)}}}
  \centerline{\includegraphics[width=0.7\linewidth]{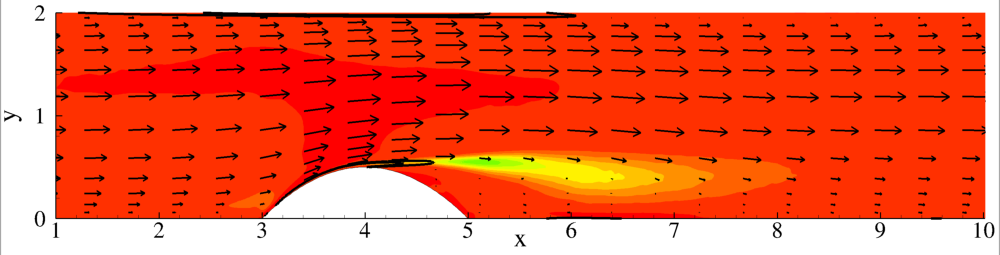}
  {\scriptsize \put(-300,63){\bf (C1)}}}
  \centerline{\includegraphics[width=0.3\linewidth]{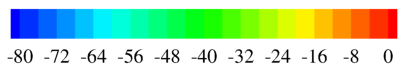}}
  \caption{Mean kinetic energy balance equation: turbulent kinetic energy production $-\Pi$ (background colour),  
  mean energy dissipation $\varepsilon_M$ (solid isolines) and mean energy spatial flux $\Phi_M$ (vectors).}
\label{fig:MKE_2}
\end{figure}

Figure~\ref{fig:MKE_2} shows the terms in equation~(\ref{eq:MKE_all}) 
normalised with the total average power injection per unit volume, i.e. the total dissipation rate, $\int \left( \varepsilon_M + \varepsilon_T \right)dV$ 
where $\varepsilon_T = \langle 1/\Rey (\partial u'_i/\partial x_j) (\partial u'_i/\partial x_j) \rangle$ is the 
turbulent dissipation rate density. In the figure, the turbulent kinetic energy production rate, $-\Pi$, 
is shown in the background colour plot whilst solid isolines (mostly concentrated near the bump wall) represent 
the mean field dissipation, $\varepsilon_M$. Vectors correspond to the spatial flux $\Phi_{Mj}$.

Given the behaviour of the mean streamwise velocity, see figure~\ref{fig:cont_u_zeroline}, the mean 
energy input, ${ U_x \Delta P}/{L_x}$, is  largest at the bump tip. On the other hand, the production $-\Pi$, 
is concentrated in the detaching shear layer well behind the bump, where the largest 
fluctuation intensities are attained, as discussed in the previous section. This region acts as a sink of mean 
energy and is fed by the mean energy flux, ${\Phi_M}_j$, that is crucial in redistributing energy from 
the external input to the turbulent production. With respect to case A1, taken as a basis for comparison, 
the maximum turbulent production increases by almost 50\% for the bluffer geometry and by 400\% 
at the maximum Reynolds number. By definition, turbulent production is the product of mean flow 
gradients and Reynolds stresses. For the given geometry, the mean gradients in the shear layer 
slightly depend on Reynolds number, as shown in panel (c) of figure~\ref{fig:uprof_mean}
which corresponds to the section of maximum production. This suggests that the mean field 
already attained an almost Reynolds independent state. On the other hand, turbulent stresses 
increase significantly at this section, see panel (c) of figure~\ref{fig:uxFuyFprof_mean}, 
resulting in the increased peak production apparent in figure~\ref{fig:MKE_2}. In general, the position 
of the energy production region depends, through the shear layer, on the dimensions and the 
position of the separation bubbles. Changing geometry at fixed, lower Reynolds number, the strength 
of the mean gradients in the same section, now in panel (d) of figure~\ref{fig:uprof_mean}, are only 
marginally affected by the change in geometry. On the other hand, the Reynolds stresses are greatly 
enhanced passing from a streamlined to a bluff configuration, panel (d) in figure~\ref{fig:uxFuyFprof_mean}, 
consistent with the increasing peak energy production from case A1 to case C1 in figure~\ref{fig:MKE_2}.

Although hardly apparent in figure~\ref{fig:MKE_2}, for the considered cases the mean flow dissipation 
rate is not irrelevant, and contributes order 40\% of the total dissipation in the system, consistently with 
significant mean velocity gradients, observed at the bump wall where the flow is abruptly accelerated.

\begin{figure}
  \centerline{\includegraphics[width=0.7\linewidth]{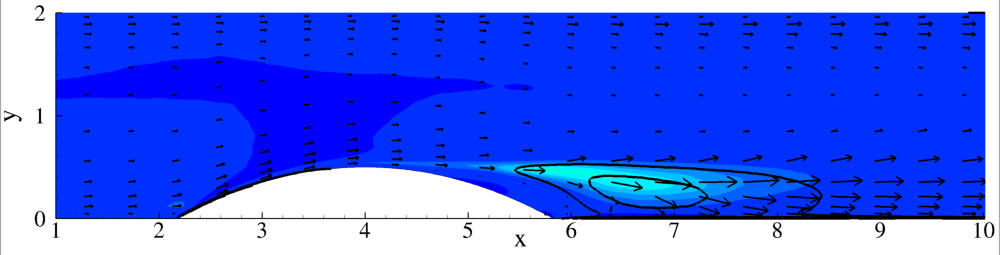}
  {\scriptsize \put(-300,63){\bf (A1)}}}             
  \centerline{\includegraphics[width=0.7\linewidth]{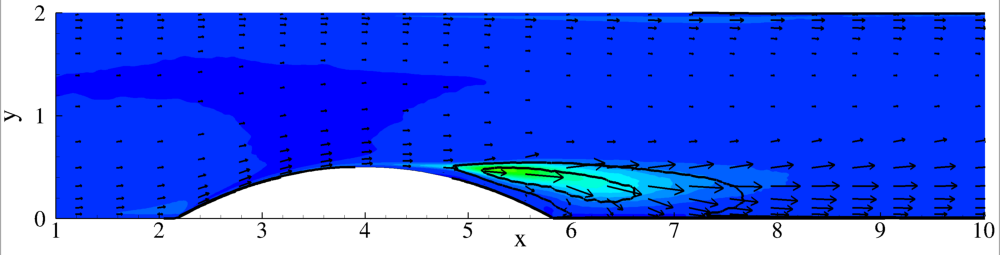}
  {\scriptsize \put(-300,63){\bf (A2)}}}            
  \centerline{\includegraphics[width=0.7\linewidth]{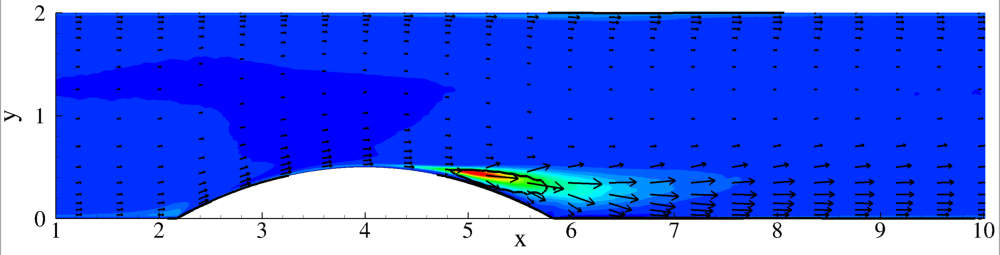}
  {\scriptsize \put(-300,63){\bf (A3)}}}           
  \centerline{\includegraphics[width=0.7\linewidth]{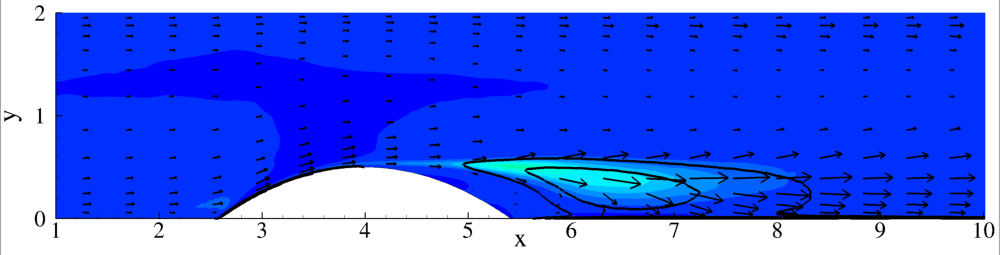}
  {\scriptsize \put(-300,63){\bf (B1)}}}          
  \centerline{\includegraphics[width=0.7\linewidth]{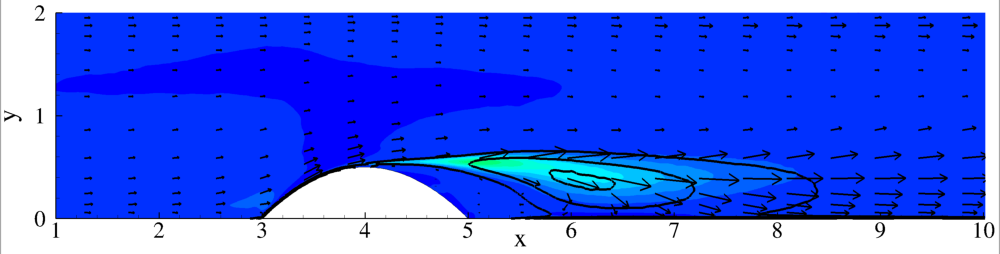}
  {\scriptsize \put(-300,63){\bf (C1)}}}         
  \centerline{\includegraphics[width=0.3\linewidth]{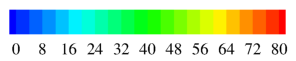}}
  \caption{Turbulent kinetic energy balance equation: turbulent kinetic energy production $\Pi$ (background colour), 
  turbulent energy dissipation $\varepsilon_M$ (solid isolines) and turbulent energy spatial flux $\Phi_M$ (vectors).}
\label{fig:TKE_2}
\end{figure}

\begin{figure}
  \centerline{\includegraphics[width=0.7\linewidth]{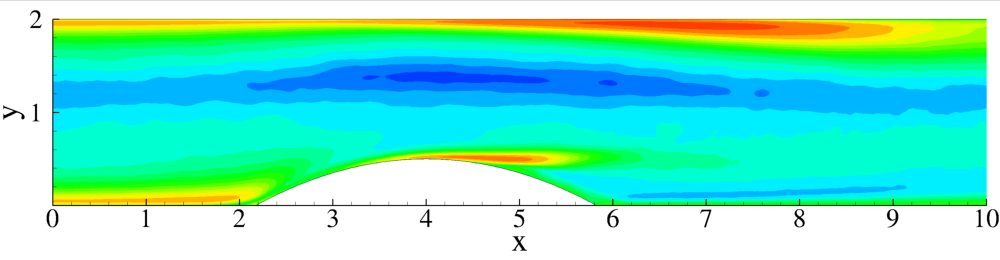}
  {\scriptsize \put(-300,63){\bf (A1)}}}                      
  \centerline{\includegraphics[width=0.7\linewidth]{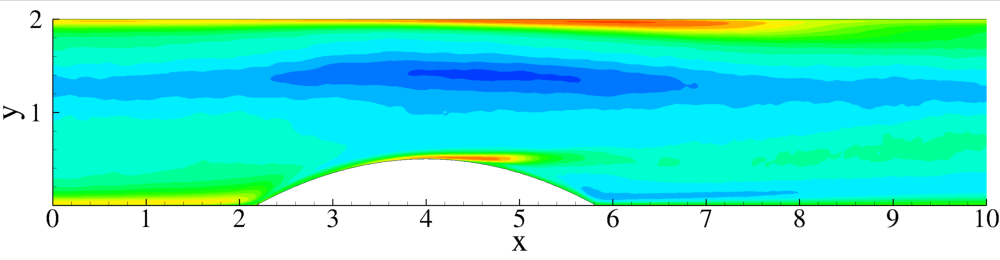}
  {\scriptsize \put(-300,63){\bf (A2)}}}                     
  \centerline{\includegraphics[width=0.7\linewidth]{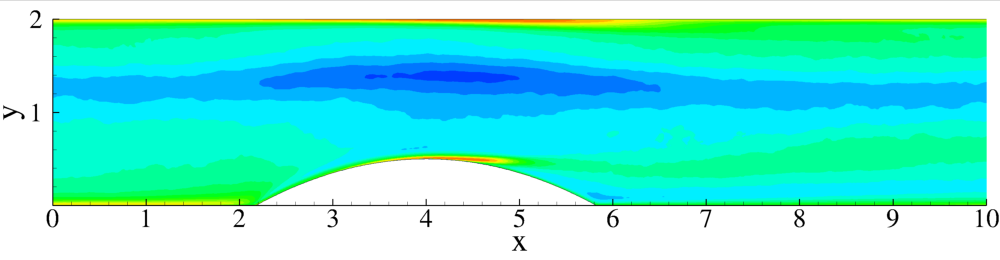}
  {\scriptsize \put(-300,63){\bf (A3)}}}                    
  \centerline{\includegraphics[width=0.7\linewidth]{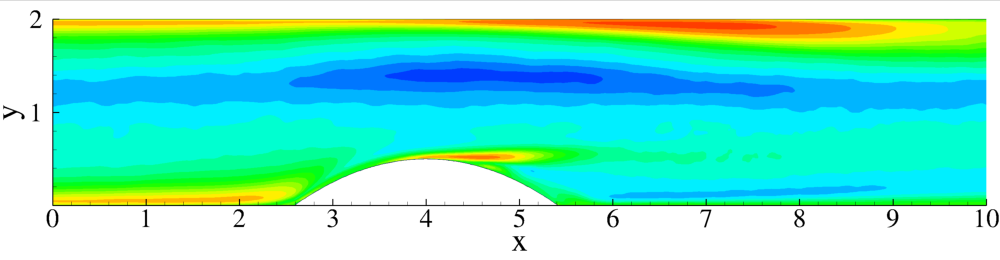}
  {\scriptsize \put(-300,63){\bf (B1)}}}                   
  \centerline{\includegraphics[width=0.7\linewidth]{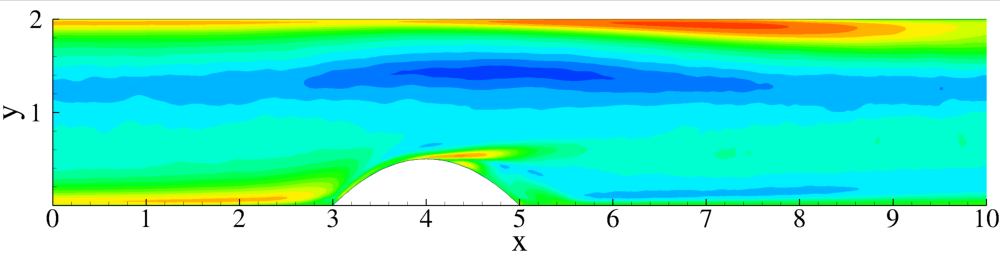}
  {\scriptsize \put(-300,63){\bf (C1)}}}                  
  \centerline{\includegraphics[width=0.3\linewidth]{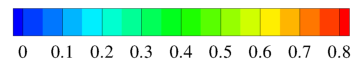}}
  \caption{Norm of the deviatoric component of the Reynolds stress tensor, $||b|| = \sqrt{b_{ij} b_{ij}}$ with $b_{ij}= \langle u'_i u'_j \rangle / \langle u'_k u'_k \rangle - 1/3\, \delta_{ij}$, and $\delta_{ij}$ the components of the identity tensor.}
\label{fig:large_scale}
\end{figure}

\subsubsection{Turbulent kinetic energy}
\label{sec:TKE}

The balance equation for the turbulent kinetic energy reads
\begin{equation}
\pd{\Phi_{Tj}}{x_j} = - \varepsilon_T + {\Pi} + \langle \frac{\Delta p'(t)}{L_x} \, u'_x \rangle\, ,
\label{eq:TKE_all}
\end{equation}
where, as anticipated, $\varepsilon_T$ is the turbulent kinetic energy dissipation rate,  $\Pi$, here with the opposite sign with respect to
eq.~\eqref{eq:MKE_all}, is the production and $\langle \Delta p' \,u'_x/L_x \rangle$  is the external source of fluctuating energy. 
The spatial flux,
\begin{equation}
\Phi_{Tj} = U_j k_T + \frac{1}{2} \langle u'_i u'_i u'_j\rangle + {\langle \tilde{p}' u'_j\rangle} - \frac{1}{\Rey}\pd{k_T}{x_j}\, ,
\label{eq:TKE_flux}
\end{equation}
contributes zero net power when integrated over the whole domain. The energy locally provided by the fluctuations 
of pressure difference between inlet and outlet $\langle \Delta p' \,u'_x/L_x \rangle$ is negligible, 
$\max\limits_{x,y}{\langle \Delta p' \,u'_x/L_x \rangle}\simeq 10^{-5} \, \max\limits_{x,y} \Pi$. 

Figure~\ref{fig:TKE_2} shows the turbulent kinetic energy production, turbulent energy dissipation and 
spatial fluxes for all the simulations. The terms are normalised with the overall power injected 
in the system which, in the statistically steady state, is balanced by the 
total dissipation rate, $\int \left( \varepsilon_M + \varepsilon_T \right)dV$.
The production term injects most  energy in the shear layer behind the bump. 
From the shear layer the energy follows different paths,  see the vector field in the figure where local energy release 
is associated with the (positive) divergence of the  energy flux. The turbulent energy  is transferred towards the centre of the channel, into the  separation bubble or towards the wall, in particular behind the  bump under the separation bubble.  
From the analysis of the dissipation field, $\varepsilon_T$, part of the energy is found to be locally dissipated in 
the shear layer and in the separation bubble. Most of the energy is dissipated at the 
bottom wall after the bump (note the isolines of dissipation concentrated in that region).

\begin{figure}
  \centerline{\includegraphics[width=0.7\linewidth]{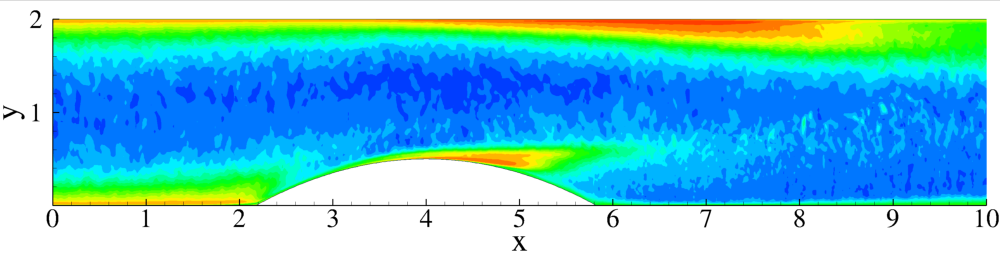}
  {\scriptsize \put(-300,63){\bf (A1)}}}                      
  \centerline{\includegraphics[width=0.7\linewidth]{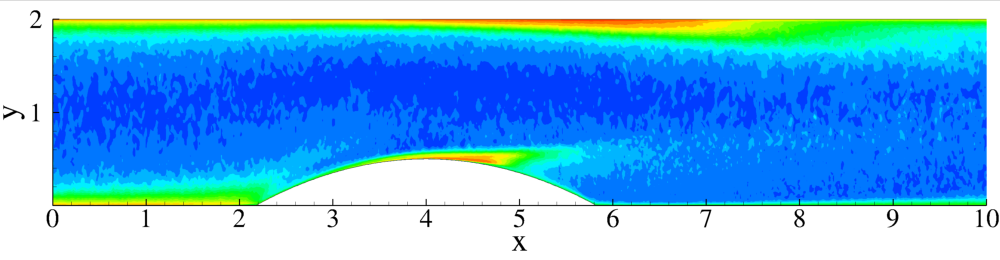}
  {\scriptsize \put(-300,63){\bf (A2)}}}                      
  \centerline{\includegraphics[width=0.7\linewidth]{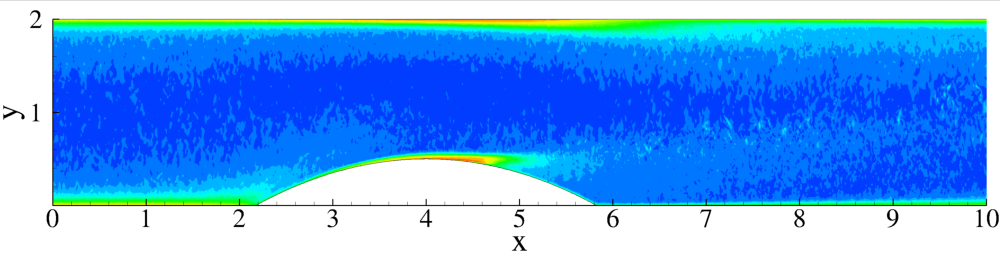}
  {\scriptsize \put(-300,63){\bf (A3)}}}                          
  \centerline{\includegraphics[width=0.7\linewidth]{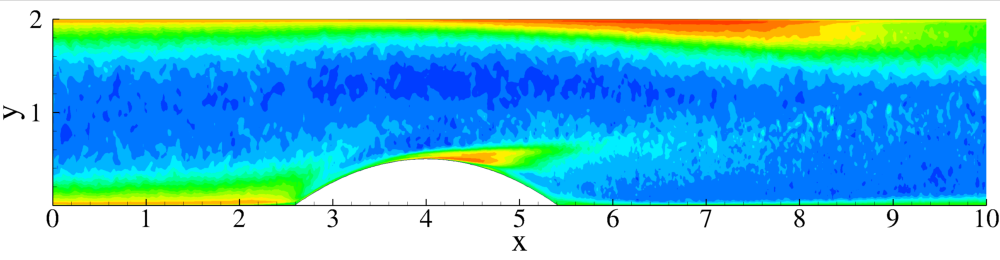}
  {\scriptsize \put(-300,63){\bf (B1)}}}                         
  \centerline{\includegraphics[width=0.7\linewidth]{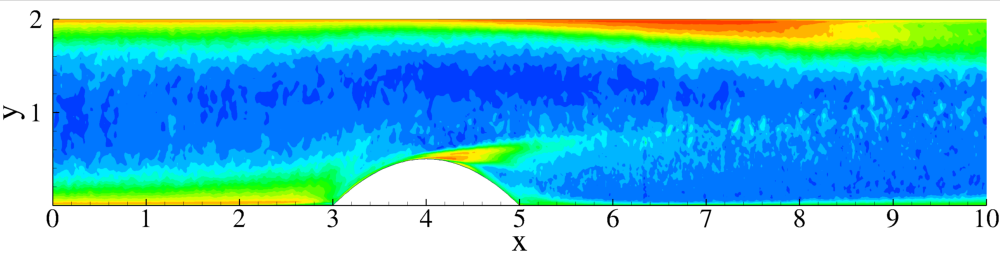}
  {\scriptsize \put(-300,63){\bf (C1)}}}                        
  \centerline{\includegraphics[width=0.3\linewidth]{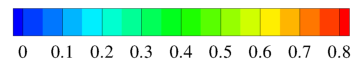}}
  \caption{Norm of the deviatoric component of the pseudo dissipation tensor, 
$||d|| = \sqrt{d_{ij} d_{ij}}$, where 
$d_{ij}=\epsilon_{ij}/\epsilon_{kk}-1/3\,\delta_{ij}$  and  
$\epsilon_{ij}=2/\Rey \langle\left(\partial u'_i/\partial x_k\right)
\left(\partial u'_j/\partial x_k\right)\rangle$ is 
  the pseudo dissipation tensor.}
\label{fig:small_scale}
\end{figure}

\subsubsection{Large and small scale anisotropy} \label{sec:ani}

The anisotropy of the large turbulent scales is described by  the deviatoric component 
of the Reynolds stress, 
$b_{ij}= \langle u'_i u'_j \rangle / \langle u'_k u'_k \rangle-1/3 \delta_{ij}$, 
where $\delta_{ij}$ denotes the Kronecker symbol. Note that in isotropic 
conditions, $b_{ij}$ is identically zero. An overall measure of anisotropy is given by 
the norm $||b|| = \sqrt{b_{ij} b_{ij}}$, figure~\ref{fig:large_scale}.  
The anisotropy is particularly significant in the near wall region and in the shear 
layer, while the bulk flow and the recirculation bubble are almost isotropic, 
consistent with the Reynolds stress profiles of figures
\ref{fig:uxF2prof_mean}, \ref{fig:uyF2prof_mean} and \ref{fig:uxFuyFprof_mean}.
Increasing the Reynolds number, the anisotropic regions become progressively 
smaller, squeezed closer to the wall, on one side, and more concentrated in the shear 
layer, on the other. The extension of the isotropic region in the bulk widens, whilst 
it shrinks with the recirculation bubble in the separated region. As a result of the 
change in geometry, the bluffest bump produces the highest anisotropic content. 

\begin{figure}
  \centerline{\includegraphics[width=0.7\linewidth]{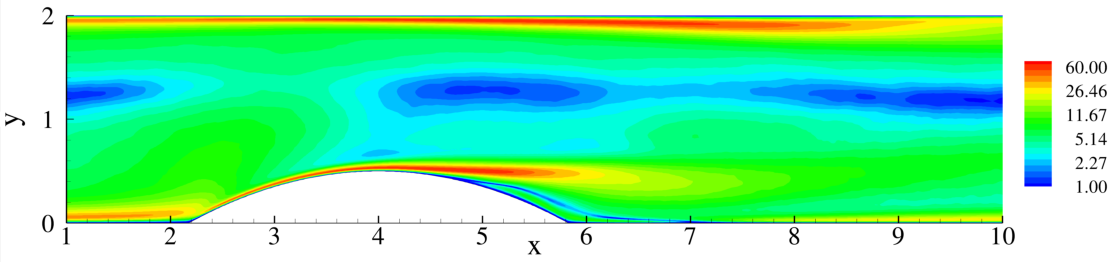}
  {\scriptsize \put(-295,60){\bf (A1)}}}
  \centerline{\includegraphics[width=0.7\linewidth]{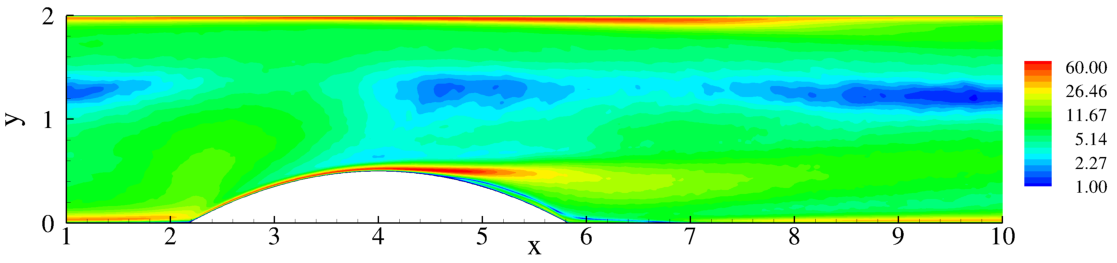}
  {\scriptsize \put(-295,60){\bf (A2)}}}
  \centerline{\includegraphics[width=0.7\linewidth]{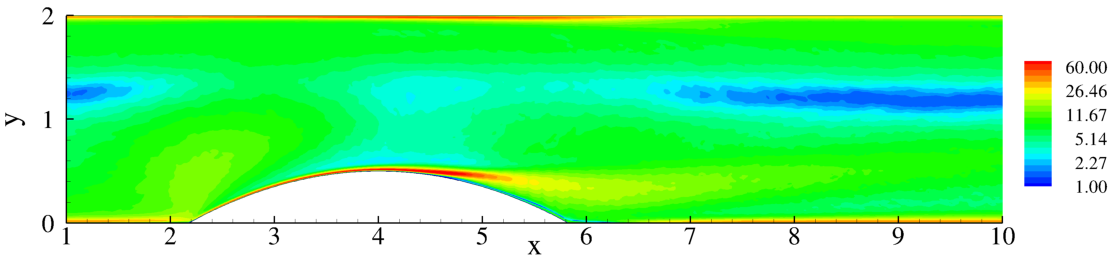}
  {\scriptsize \put(-295,60){\bf (A3)}}}
  \centerline{\includegraphics[width=0.7\linewidth]{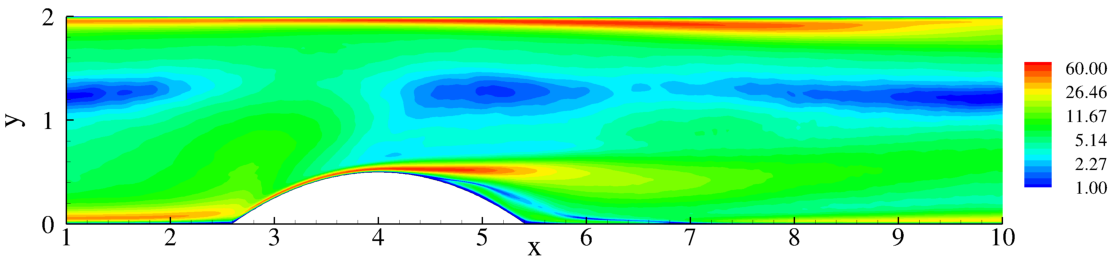}
  {\scriptsize \put(-295,60){\bf (B1)}}}
  \centerline{\includegraphics[width=0.7\linewidth]{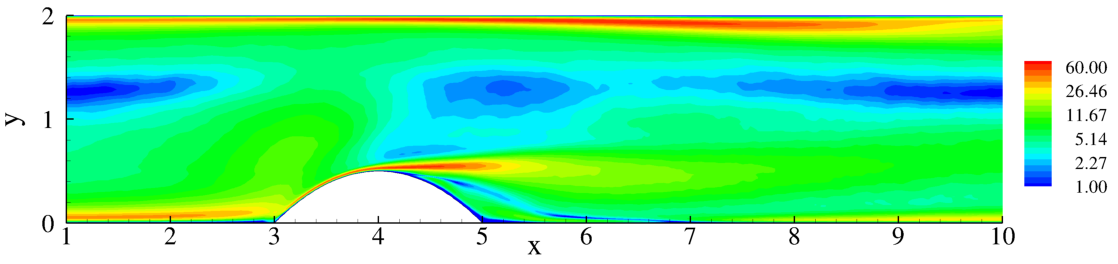}
  {\scriptsize \put(-295,60){\bf (C1)}}}
  \caption{Shear intensity 
   $S^*=S\, (2 k_T)/\epsilon_T=\left(L_0/L_S\right)^{2/3}$ in the flow domain for all simulations. }
\label{fig:S_star}
\end{figure}

\begin{figure}
  \centerline{\includegraphics[width=0.7\linewidth]{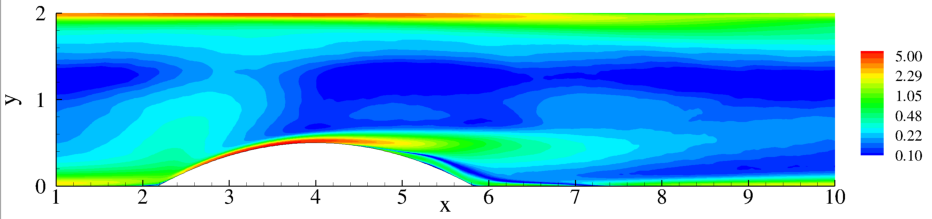}
  {\scriptsize \put(-295,60){\bf (A1)}}}           
  \centerline{\includegraphics[width=0.7\linewidth]{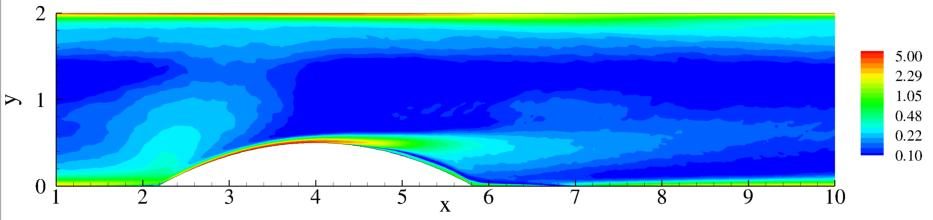}
  {\scriptsize \put(-295,60){\bf (A2)}}}          
  \centerline{\includegraphics[width=0.7\linewidth]{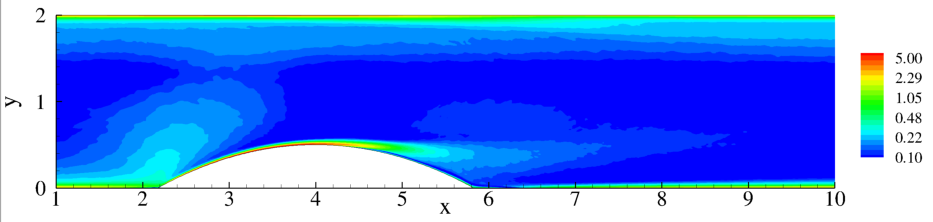}
  {\scriptsize \put(-295,60){\bf (A3)}}}         
  \centerline{\includegraphics[width=0.7\linewidth]{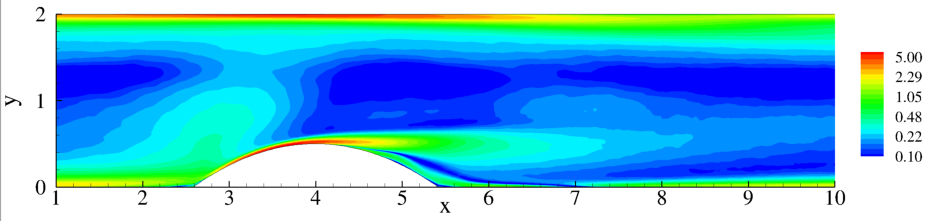}
  {\scriptsize \put(-295,60){\bf (B1)}}}        
  \centerline{\includegraphics[width=0.7\linewidth]{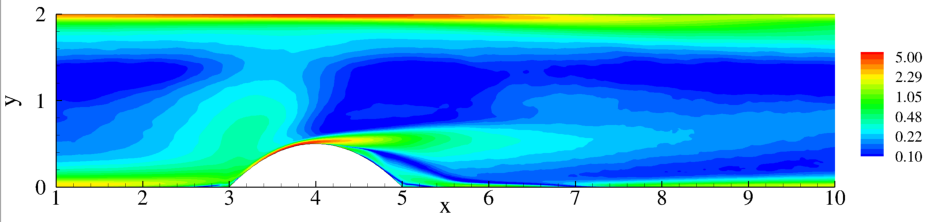}
  {\scriptsize \put(-295,60){\bf (C1)}}}
  \caption{Corrsin parameter 
  $S_c=S\sqrt{\nu/\epsilon_T}=\left(\eta/L_S\right)^{2/3}$ in the flow domain for all simulations.}
\label{fig:S_c}
\end{figure}

Figure~\ref{fig:small_scale} reports the norm, $||d||=\sqrt{d_{ij}d_{ij}}$, of the 
deviatoric component, $d_{ij} = \epsilon_{ij}/\epsilon_{kk} - 1/3\, \delta_{ij}$, 
of the pseudo-dissipation tensor, 
$\epsilon_{ij}=2/\Rey \langle(\partial u'_i/\partial x_k)(\partial u'_j/\partial x_k)\rangle$. 
$||d||$ provides a measure of the small scale anisotropy content
\citep{antonia1994anisotropy,pumir2016small} and therefore, as $||d||$  approaches zero, 
isotropic behaviour of the smallest scales is achieved. 
The small scales in the recirculation bubble and in the bulk of the flow are isotropic, 
consistently with the isotropy of the large scales in the same regions. Strong anisotropy 
persists in the near wall  regions and in the shear layer. The behaviour of $||d||$ is 
strongly dependent on the Reynolds number which ultimately sets the separation between 
the largest and the smallest scales. The regions of small-scale isotropy progressively 
increase with the Reynolds number, basically due to the shrinking of the large scale 
anisotropy regions. However, anisotropy still persists at small scales, irrespectively of 
the Reynolds number, near the walls and in the shear layer. 
This behaviour can be explained and understood by addressing the
dynamics of a turbulent flow in presence of strong shear.
In isotropic conditions the turbulence is forced at the largest scales
comparable with the integral scale $L_0=(2 k_T)^{3/2}/\epsilon_T$ and
is dissipated by viscosity at the Kolmogorov scale $\eta = (\nu^3/\epsilon_T)^{1/4}$.
In the inertial range ($\eta \ll r \ll L_0$) the energy is simply transferred from 
the large to the small scales. In turbulent shear flows the shear scale $L_S=\sqrt{\epsilon_T/S^3}$, extensively discussed  in~\cite{Casciola_2003}, where $S$ is the shear rate, plays a crucial role in explaining the dynamics.  Basically the shear scale identifies the range of scales $L_S < r < L_0$ where the
turbulence is driven by the (anisotropic) production of turbulent kinetic energy
due to the Reynolds stresses. In the range of scales below $L_S$,  
$\eta < r < \L_S$, the dynamics of the turbulent fluctuations are driven by the process 
of energy cascade typical of isotropic flows, see e.g.~\cite{Marati_2004,cimarelli2013paths} 
for a detailed analysis of the energy paths in a planar channel.
It follows that the dynamics of a shear flow is described by
two dimensionless  parameters. The first one is the shear intensity 
$S^*=S\, (2 k_T)/\epsilon_T$, see e.g.~\cite{lee1990structure}, that can be recast in terms
of the shear scale as $S^*=\left(L_0/L_S\right)^{2/3}$,~\citep{casciola2007residual}.
The shear intensity measures the separation between the shear scale and the integral scale
thus providing the extension of the range of scales directly affected by the
production mechanisms. The second parameter is the Corrsin parameter 
$S_c=S\sqrt{\nu/\epsilon_T}$ \citep{corrsin1958local} that can be recast 
in terms of the shear scale as $S_c=\left(\eta/L_S\right)^{2/3}$. The Corrsin 
parameter measures the extension of the range between the shear scale and the Kolmogorov scale
where the flow is driven by the inertial cascade. Clearly, only in the
range of scales below the shear scale isotropisation of turbulent fluctuations
can take place. The shear scale can be evaluated in spatially non 
homogeneous flows by considering the norm of the local mean velocity gradient
$L_S = \sqrt{\epsilon_T/\left(\partial_j U_i \partial_j U_i\right)^{3/2}}$. 
In our case the shear scale is a field $L_S(x,y)$. In a similar way the
local integral scale $L_0=(2 k_T)^{3/2}/\epsilon_T$ and the local Kolmogorov
scale $\eta = (\nu^3/\epsilon_T)^{1/4}$ can be considered. The shear strength
$S^*$ and the Corrsin parameter $S_c$ are position dependent.
When $S^*$ is large the whole range of scales is dominated by 
production and there is no room left for isotropy recovery at small scales.
On the contrary, an isotropy recovery range is available where the
Corrsin parameter is small.
Figures \ref{fig:S_star} and \ref{fig:S_c} provide the fields  $S^*$ and
$S_c$ respectively, for the different Reynolds numbers and geometries considered
in this paper. A joint analysis of $S^*$ and $S_c$  provides the physical interpretation of 
the observed anisotropy (figures~\ref{fig:large_scale} and~\ref{fig:small_scale}). 
In the bulk region, $S^*$ decreases, denoting weak production of turbulent kinetic
energy since the shear scale approaches the integral scale. Concurrently, $S_c$ is small, indicating 
a large separation between shear and Kolmogorov scale. This behaviour is generic and the only relevant changes are
observed when the Reynolds number is increased, A1-A3. At large Reynolds number, 
the spatial region in the bulk where isotropisation occurs is broadened.
The relative position of integral, shear and Kolmogorov scales in the 
bulk explains why the small scales are isotropic (figure~\ref{fig:small_scale}).
The conditions are different near the wall and in the shear layer. 
In these regions,  see figure~\ref{fig:S_star}, $S^*$ is large and the whole range of scales is now dominated by 
turbulent kinetic energy production. Concurrently $S_c$ is  order one, i.e. the shear scale is forced on the 
Kolmogorov scale (figure~\ref{fig:S_c}). This behaviour is again generic for 
the cases we address. In a nutshell, near the
wall and in the shear layer there is no room for the formation of the inertial range
where isotropisation can take place. The flow is driven by the anisotropic
mechanisms of turbulent kinetic energy production, see figure~\ref{fig:large_scale}
and the anisotropy persists down to the smallest scales (figure~\ref{fig:small_scale}).

%%%%%%%%%%%%%%%%%%%%%%%%%%%%%%%%%%%%%%%%%%%%%%%%%%%

\subsubsection{Invariant Maps} \label{sec:inv_maps}

The Anisotropy Invariant Map (AIM) originally introduced by~\cite{lumley1977return,lumley1979computational} provides a description of the different anisotropic states of the large turbulent scales.
They are quantified in terms of the invariants of the
anisotropy tensor, i.e. the deviatoric component of the Reynolds stress, $b_{ij}$, 
namely $I = b_{ii}=0$, $II=b_{ij}\, b_{ji}$ and  $III=b_{ij} \, b_{jk} \, b_{ki}$.  
The admissible states of the flow must lie  within a (curvilinear) triangle of the $II - III$ plane. 
This constraint comes from  the requirement that the eigenvalues of $b_{ij}$ should be real 
and the squared velocity fluctuation in the principal direction must be positive.
The admissible region is delimited above by the line $II=2/9 + 2 III$ corresponding to statistically 
two-dimensional turbulence, i.e. the fluctuation intensity in one of the eigen-directions vanishes.
The other two limiting lines, $II=3/2 \left( 16/9 \, III^2\right)^{1/3}$, represent axisymmetric 
turbulence, i.e.  the fluctuation intensity in two eigen-directions are identical.
In the left branch ($III<0$), the fluctuation intensity in the third eigen-direction is smaller than 
the other two (pancake turbulence). In the right branch ($III>0$), the third  component is larger 
than the other two (cigar-like turbulence). The corners correspond
to: the isotropic state ($II=0, \, III=0$),
the two-component isotropic state ($II=-1/36, \, III=1/6$), and the one-component state 
($II=2/9, \, III=2/3$).
In many applications, turbulence modelling exploits the idea of eddy viscosity
which assumes that the Reynolds stress tensor is proportional (via the eddy-viscosity)
to the mean strain rate tensor, see~\cite{speziale1991analytical,gatski1993explicit}.
Modelling is particularly challenging for separated flows. The AIM is
helpful to directly take into account the anisotropy of the flow,
see~\cite{jovivcic2006anisotropy,kumar2009anisotropy} and the general discussion
in~\cite{jovanovic2013statistical} where $II$ and $III$ are used to 
compute the length scale appearing in the eddy-viscosity thus including anisotropic 
effects in the model.
\begin{figure}
  \centerline{
  \includegraphics[width=0.32\textwidth]{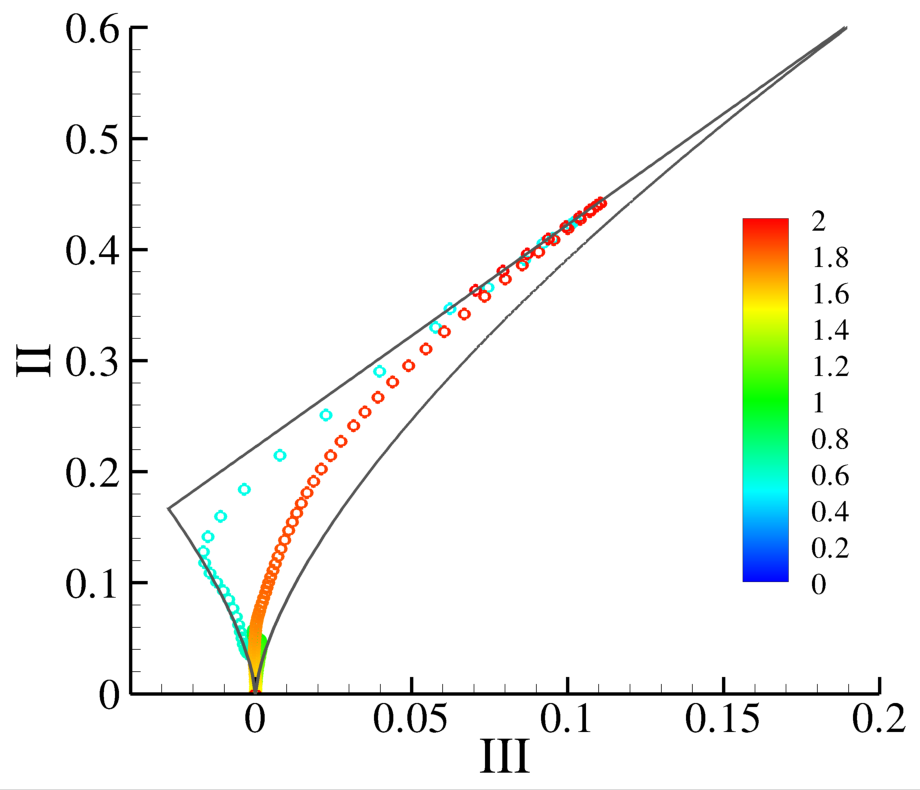}
    {\scriptsize \put(-130,100){\bf (a)}}
  \includegraphics[width=0.32\textwidth]{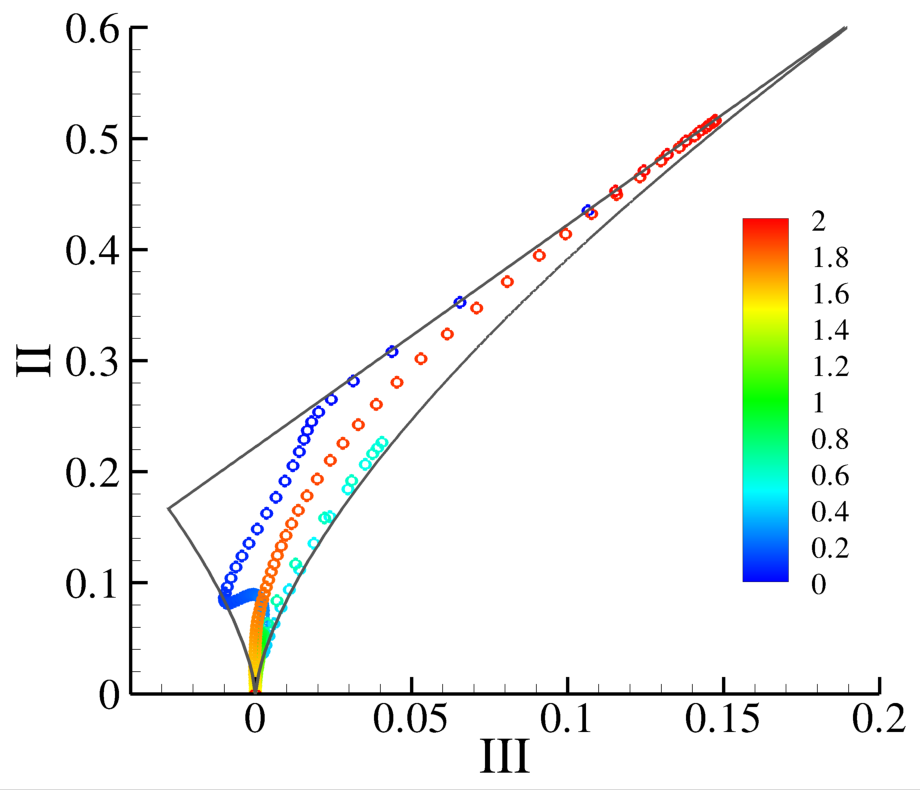}
    {\scriptsize \put(-130,100){\bf (b)}}
  \includegraphics[width=0.32\textwidth]{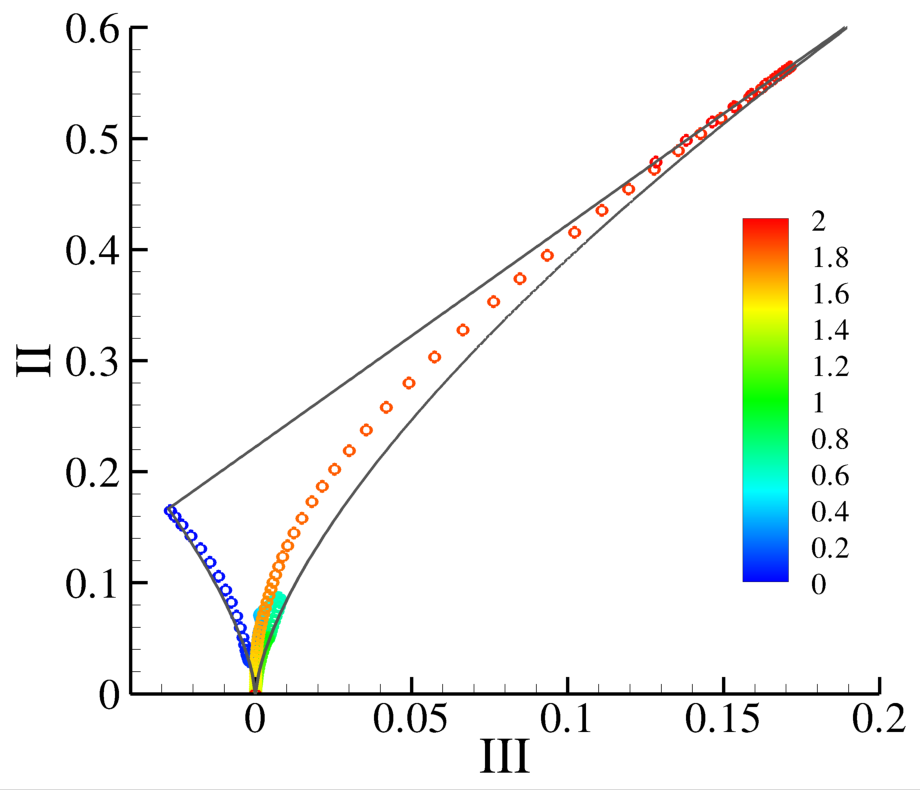}
    {\scriptsize \put(-130,100){\bf (c)}}}
  \caption{Anisotropy Invariant Map for case C1 at: (a) the tip of the bump;
(b) end of the bump; (c) inside the recirculation region. Note that (a), (b) and (c) correspond
to stations b), c) and d) of figure\ref{fig:stations}. The color legend represents
the $y$ coordinate.} \label{fig:inv_maps_C1}
\end{figure}

Figure~\ref{fig:inv_maps_C1} shows the AIM for simulation C1 at three stations. At the tip of the bump, panel (a),
close to the bump wall at $y=0.5$, turbulence is essentially two-dimensional. As $y$
increases, the points in the plot approach the lower left branch indicating axisymmetric
turbulence. As expected, close to the centreline of the channel, the flow becomes isotropic. As the top
wall is approached, the flow follows a trend similar to what is found in a planar channel
\citep{gilbert1991turbulence} becoming two-dimensional again (red points) at the top 
wall. However, in our case due to the adverse pressure gradient at the top wall the trajectory
followed in the map by the orange/red points is shifted away from the right branch,
departing from the axisymmetric state typical of the planar channel flow.
Panel (b) corresponds to the end of the bump which exhibits a more complex AIM. As the 
$y$ coordinate increases, the initial axisymmetric state is hardly reached 
and the trajectory in the $II-III$ plane follows an inner path towards the opposite
axisymmetric state ($III>0$) as the shear layer is crossed (light blue). At the centreline, the
flow is again isotropic and follows an inner path towards the top wall. These 
results are typical of separated flows as found, e.g., in the backward facing step configuration
\citep{le1997direct}. Panel (c) shows the
trajectory at a station closer to the reattachment point. Flow close to the lower wall 
is completely axisymmetric ($III<0$) until the shear layer is reached (light blue).
The turbulence shifts from axisymmetric contraction to axisymmetric expansion, similar to panel (b).
The flow is isotropic at the centreline and follows the same trend discussed for panel (a) and 
(b) as the top wall is reached. The analysis of the AIM suggests that the flow we are addressing 
is rather complex to model, due to the presence of the recirculating region behind the 
bump and the adverse pressure gradient along the top wall.

\section{Conclusions}
\label{sec:final}

Turbulent separation behind a bump in channel flow is addressed using Direct Numerical 
Simulations (DNS) for different bump geometries and for Reynolds number ranging 
between $\Rey=2500$ and $\Rey=10000$. The latter is probably the largest Reynolds number 
ever achieved in the DNS of this specific configuration, corresponding to a maximum 
friction Reynolds number of approximately $\Rey_\tau=900$. 

The separation behind the bump generates small scale structures which grow downstream, 
an intense shear layer and a recirculating region after the bump. Although the recirculation 
size depends on geometry, the reattachment position is constant.
The reattachment point is controlled  by the Reynolds number based on the bump height, 
as confirmed by the available experimental data, \cite{Kahler_2016}.

With increasing Reynolds number, a net decrease in drag coefficient is observed in 
association with the reduced dimensionless pressure drop needed to maintain the 
flow rate constant. The reduction is overwhelmed by the increase in the dimensional 
velocity, quadratically entering the expression for the drag force, leading to 
the expected increase in flow resistance. The drag increase with respect to that of 
an equivalent planar channel is almost entirely due to form effects induced by the 
separation, even though a significant increase in velocity, hence in local wall shear stress,
is measured at the bump tip. At larger Reynolds number, the shear layer separating the 
recirculation bubble from the outer stream  becomes more attached to the lower wall. 
Its fluctuations correspond to higher turbulent kinetic energy production peaks. 
The DNS captures a small recirculation originated by the sudden change in slope at the 
bump leading edge. The separation at the bottom wall affects the opposite near-wall region 
by inducing a significant adverse pressure gradient which is not sufficient to separate 
the flow at the upper wall. 

Due to the strong non-homogeneity and the resulting mean gradients, the mean flow draws energy 
from the local external energy source, namely the pressure drop multiplied by velocity. 
The uptake mostly occurs in the bulk. Fluxes move this energy to the shear layer 
where it is partially dissipated but mostly intercepted by the production term to
sustain the turbulent fluctuations. The dissipation in the mean flow is significant, 
given the strong mean gradients present at the walls and in the shear layer. Overall, 
the most important feature is the peak of turbulent kinetic energy production localised in 
the shear layer. The path taken by this energy bifurcates, in part sustaining  the turbulent 
fluctuations inside the bubble and in part feeding the turbulence of the external flow 
downstream of the bubble. From the most streamlined to the bluffest bump, a 50\%  
increase in peak energy production is observed. For fixed geometry, a fourfold Reynolds 
number change leads to approximately 400\% increase in peak production.

In turbulence modelling, the level of anisotropy at both the large and small scales is crucial. 
They can be characterised in terms of the deviatoric components of Reynolds stress and 
pseudo-dissipation tensor, respectively. Apart from the near wall region, anisotropy at both 
large and small scales concentrates in the shear  layer, irrespective of bump shape and 
Reynolds  number.  Interestingly, the small scales keep a high level of anisotropy in the 
shear layer, even at larger Reynolds number. This is due to the intensity of 
the mean gradients which maintain the production active close to dissipation scales.
Finally, the analysis of the anisotropy invariant maps shows that the separated flow 
poses a significant difficulty for turbulence modelling due to the recirculating region behind 
the bump and the adverse pressure gradient along the top wall.

%Acknowledgements
\section*{}
The research has received funding from the European Research Council under the ERC Grant Agreement no. 339446. 
We acknowledge PRACE for awarding us access to supercomputing resource FERMI based in Bologna, Italy through PRACE project no. 2014112647.

\bibliographystyle{jfm}
\bibliography{biblio}

\end{document}